\documentclass[english,pra,aps,superscriptaddress,notitlepage,nofootinbib,mathrsfs,twocolumn]{revtex4-1}

\usepackage{color}
\usepackage{mathtools}
\usepackage{amsmath}
\usepackage{amssymb}
\usepackage{graphicx}

\usepackage{bm}

\usepackage{babel}
\usepackage[colorlinks=true, allcolors=blue]{hyperref}

\newcommand{\btxt}[1]{{\color{black} #1}}

\begin{document}

\title{Floquet theory for temporal correlations and spectra in time-periodic\\ open quantum systems: Application to squeezed parametric oscillation \\beyond the rotating-wave approximation}

\author{C. Navarrete-Benlloch}
\email{derekkorg@gmail.com}

\affiliation{Wilczek Quantum Center, School of Physics and Astronomy, Shanghai
Jiao Tong University, Shanghai 200240, China}

\affiliation{Max-Planck-Institut f\"{u}r die Physik des Lichts, Staudtstra{\ss}e 2,
91058 Erlangen, Germany}

\affiliation{Shanghai Research Center for Quantum Sciences, Shanghai 201315, China}

\author{R. Garc\'{e}s}

\affiliation{Departament d'\`{O}ptica i Optometria i Ci\`{e}ncies de la Visi\'{o}, Facultat
de F\'{i}sica, Universitat de Val\`{e}ncia, Dr. Moliner 50, 46100 Burjassot,
Spain}

\author{N. Mohseni}

\affiliation{Max-Planck-Institut f\"{u}r die Physik des Lichts, Staudtstra{\ss}e 2,
91058 Erlangen, Germany}

\author{G. J. de Valc\'{a}rcel}
\email{german.valcarcel@uv.es}

\selectlanguage{english}%

\affiliation{Departament d'\`{O}ptica i Optometria i Ci\`{e}ncies de la Visi\'{o}, Facultat
de F\'{i}sica, Universitat de Val\`{e}ncia, Dr. Moliner 50, 46100 Burjassot,
Spain}

\begin{abstract}
Open quantum systems can display periodic dynamics at the classical
level either due to external periodic modulations or to self-pulsing phenomena typically following a Hopf bifurcation. In both cases, the quantum fluctuations around classical solutions do not reach a quantum-statistical stationary state, which prevents adopting the simple and reliable methods used for stationary quantum systems. Here we put forward a general and efficient method to compute two-time correlations and corresponding spectral densities of time-periodic open quantum systems within the usual linearized (Gaussian) approximation for their dynamics. Using Floquet theory we show how the quantum Langevin equations for the fluctuations can be efficiently integrated by partitioning the time domain into one-period duration intervals, and relating the properties of each period to the first one. Spectral densities, like squeezing spectra, are computed similarly, now in a two-dimensional temporal domain that is treated as a chessboard with one-period $\times$ one-period cells. This technique avoids cumulative numerical errors as well as efficiently saves computational time. As an illustration of the method, we analyze the quantum fluctuations of a damped parametrically-driven oscillator (degenerate parametric oscillator) below threshold and far away from rotating-wave approximation conditions, which is a relevant scenario for modern low-frequency quantum oscillators. Our method reveals that the squeezing properties of such devices are quite robust against the amplitude of the modulation or the low quality of the oscillator, although optimal squeezing can appear for parameters that are far from the ones predicted within the rotating-wave approximation.
\end{abstract}
\maketitle

\section{Introduction}

In recent years, with the development of new quantum technologies,
more complex protocols to control and manipulate quantum devices have
been proposed. Usually, those devices are made up of nearly isolated
quantum systems (atoms, solid-state defects, superconducting circuits,
mechanical elements, etc) that interact coherently with the electromagnetic
field (at optical or microwave frequencies) via an input port, where
a driving is applied, and an output port, where the detection is performed.
Also, the considered quantum system can interact with its environment,
usually leading to an incoherent exchange of excitations which manifests
as noise (thermal, electronic, etc). A proper engineering of all these
processes is key to design quantum technologies for applications in
quantum computation, simulation, communication, and metrology.

In the context of cavity quantum optics (where we also consider the
related fields of superconducting-circuit resonators, polariton microcavites,
and cavity optomechanics), as well as in the field of many-body physics,
several works based on periodically-modulated driving have appeared
recently in the literature aimed at controlling or enhancing specific
features, as well as promoting the emergence of new phenomena and
even novel phases of matter. The use of modulations has been proposed,
for instance, for generating two-mode entangled states in superconducting
circuit resonators \citep{QubitEntangled}, quantum squeezing of the
mirror motion \citep{MechSqueezingTh,MechSqueezing,Mari,Farace} or
of the radiation field in optomechanical \citep{LightSqueezing,Garces2016,Levitan}
and superconducting-circuit cavities \citep{Garces2016}, for producing
entanglement between a mechanical and an optical mode or between two
radiation modes \citep{Farace,Mari-NJP,Liao}, for entangling the
motional degrees of freedom of two tethered and optically-trapped
microdisks inside a cavity \citep{Abdi}, for cooling the ground state
of a mechanical oscillator \citep{rocking-Zhang}, for measuring the
position of a mechanical oscillator in an optomechanical backaction-evading
scheme \citep{Evasion,Evasion1}, for enhancing nonlinear interactions
in quantum optomechanics \citep{Lemond}, or for synchronization or
entrainment purposes \citep{sync1,sync2,sync3,sync4,sync5,sync6,sync7,sync8,sync9,sync10}
with its implications in the emergence of quantum correlations and
entanglement \citep{Zambrini,Timme}. Such periodic or multi-periodic
drivings can also be used to engineer elusive dissipative models such
as squeezed lasers \citep{CNB-SqueezedLasing}, degenerate parametric
oscillation \citep{CNB-DPO-OM,Legtas15}, and non-reciprocal devices
\citep{Anja2,Anja3,Anja4,Anja5} with a range of applications \citep{AnjaApp1,AnjaApp2,Anja1}.
Moreover, spontaneous periodic oscillations (also called limit cycles)
can emerge in nonlinear systems, usually via Hopf bifurcations. In
particular, such oscillations have been observed experimentally in
optomechanical cavities operating in the classical regime \citep{HBexp1,HBexp2,HBexp3,HBexp4},
and are well understood theoretically \citep{HBth1,HBth2,HBth3}.
In contrast, the study of quantum dynamics around limit cycles is
a very active field of theoretical research in different contexts
\citep{HBq1,HBq2,HBq4,HBq5,HBq6,LC,LC-Hartmann,LC-JJ,RefA1,RefA2,RefA3}, including
their connection to the emergence of time crystals \citep{Frank12quantum,Frank12classical,Frank19,TCreview}
in driven-dissipative many-body systems \citep{OpenTC1,OpenTC2,OpenTC3,OpenTC4,OpenTC5,OpenTC5bis,OpenTC6,OpenTC7}.
Closely related to the latter is the field of Floquet or discrete
time crystals in periodically-driven closed many-body systems \citep{DTC1,DTC2,DTC3,TCreview,DTCapps1},
which have only recently been identified as unconventional phases
of matter far from equilibrium \citep{DTC4,DTC5,DTC6,DTC7,DTC8},
but have already sparked interesting experiments \citep{DTCexp1,DTCexp2,DTCexp3,DTCexp4,DTCexp5,DTCexp6,DTCexp7,DTCexp8}
and applications \citep{DTCapps1,DTCapps2}. Also in this context,
periodic modulations allow for the so-called Floquet engineering of
Hamiltonians \citep{Shirley}, leading to some desired properties such as nontrivial topology or optimized transport \citep{Hanggi98,FloquetEngineering}.

Taking all these things into consideration, it is clear that periodic
modulations play a major role in many different fields of contemporary
quantum physics. In this work, we will concentrate on periodically-driven
open quantum-optical systems. There are two standard mathematical
descriptions of such quantum systems: (i) via a set of coupled quantum
Langevin equations, which are Heisenberg (differential) equations
for the operators supplemented by dissipation terms and input quantum
noises, or (ii) via a master equation for the density operator, which
consists on the von Neumann equation for the state, to which Lindblad
terms accounting for irreversible quantum jumps are added. Let us
remark that master equations can be mapped to a set of stochastic
Langevin equations by resorting to phase-space representations of
the density operator (like the Wigner function or, more robustly,
the positive \emph{P} distribution \citep{PositiveP1,PositiveP2,PositiveP3}).
Hence, in both approaches a set of Langevin equations can be ultimately
obtained, which provide a route towards the numerical analysis of dynamical features.

Due to the generally nonlinear nature of such equations, exact solutions
are hard or impossible to find, except for very specific cases. Analytic
or semi-analytic insight is usually gained by using the so-called
standard linearization technique, which typically provides sensible
results, except close to phase transitions or in the presence of spontaneous
symmetry breaking, which nevertheless can be treated with suitable
generalizations of such technique \citep{CNB08,CNB09,CNB10,CNB-PhD,CNB14-Linear,CNB15-Linear,SPOPOrot,LC}.
Within this approach, one considers small quantum fluctuations around
a reference classical state, leading to a linear system of Langevin
equations for the fluctuations, which is easily handled only if the
classical reference is time independent. This can occur in problems
involving a constant pump, like the laser, or even in the presence
of a monochromatic drive, as in optical parametric oscillators \citep{CNB-PhD}
and optomechanical cavities \citep{OM} (but only if a rotating-wave
approximation can be invoked). However, even in such cases, the stationary
classical solutions can become unstable (e.g. via a Hopf bifurcation),
and spontaneous oscillations can emerge in the classical dynamics,
leading to nontrivial linearized Langevin equations for the fluctuations,
which in particular will contain now time-periodic coefficients that
make them hard to treat. The same happens if the drive contains more
than one frequency (or if the rotating-wave approximation cannot be
used), in which case it is in general impossible to obtain time-independent
classical states.

For this type of linear Langevin equations with time-periodic coefficients,
common strategies are based on Fourier expansions \citep{Mari,FloquetApproach,FourierApproach}.
Recently, however, we put forward a more compact approach based on
the Floquet theorem \citep{LC}, which transforms linear homogeneous
differential equations with periodic coefficients into equivalent
equations with constant coefficients. There, however, we focused on
the determination of the asymptotic covariance matrix for long times,
and its connection with the steady state of the master equation after
diffusion around the limit cycle has taken over. In this work we go
deeper into the general Floquet method for linearized systems, in
particular using it to develop an efficient method for the computation
of experimentally-relevant quantities such as two-time correlation
functions and the corresponding spectral densities. Specifically,
we show how these quantities can be evaluated just from knowledge
of the behavior of the system during a single period, which
is crucial in order to avoid significant errors in the computation
of such observables: a large numerical effort can be employed at a low cost to perform highly precise integrations along one period, and then propagate that information algebraically over the long term. 

As a practical example, we use the theory to
analyze the squeezing properties of degenerate parametric oscillators
beyond the rotating-wave approximation, which has become a timely
issue, since such a model can be implemented nowadays in low-frequency
superconducting oscillators well within the quantum regime \citep{Legtas15}.
Our results support the robustness of squeezing against the modulation
amplitude or the bad quality of the oscillator. Moreover, we show
that once counter-rotating terms are incorporated, optimal squeezing
is achieved for modulation amplitudes below the oscillation instability,
contrary to the rotating-wave predictions, for which optimal squeezing
always occurs at the instability. 

The manuscript is organized as follows. In Sec. II we briefly review
the description of open quantum systems via linearized Langevin equations,
and introduce the Floquet-based method for the determination of their
solutions. In Secs. III-V we use the solutions to manipulate two-time
correlations and the corresponding spectral densities, producing compact
expressions solely based on dynamics over a single period. Finally,
in Sec. VI we apply the theory to degenerate parametric oscillation
beyond the rotating-wave approximation.

\section{Linearization in time-periodic open quantum systems: Floquet theory\label{sec:2}}

Consider an open quantum system furnished with a set of $D$ operators
$\hat{\boldsymbol{r}}=(\hat{r}_{1},...,\hat{r}_{D})^{\mathsf{T}}$,
where the symbol $\mathsf{T}$ denotes transposition. The system evolves
according to its own dynamics as well as to interactions with its
environment, which in general is composed of several reservoirs with
which the system exchanges energy. In the Heisenberg picture, which
we adopt, and assuming standard Markovian conditions, the system operators
evolve generically according to some quantum Langevin equations
\begin{equation}
\frac{d\hat{\boldsymbol{r}}}{dt}=\mathbf{A}(\lambda;\hat{\boldsymbol{r}})+\mathcal{B}(\lambda;\hat{\boldsymbol{r}})\hat{\boldsymbol{\xi}}(t).\label{H-L}
\end{equation}
Here $\mathbf{A}=(A_{1},...,A_{D})^{\mathsf{T}}$ accounts for the
deterministic, Hamiltonian or not, part of the dynamics and depends on the
system operators and also on a set of control parameters generically
denoted by $\lambda$ (e.g. the amplitude and frequency of a driving
field). These can be time dependent thereby inducing
periodic dynamics in the classical limit. The fluctuations fed by
the reservoirs into the system, responsible for irreversible quantum
jumps, enter the dynamics through a noise term. With full generality,
we write it as a $(D\times N)$ matrix $\mathcal{B}$ (that might
depend as well on the control parameters and the system operators)
acting on a vector $\hat{\boldsymbol{\xi}}(t)=(\hat{\xi}_{1}(t),...,\hat{\xi}_{N}(t))^{\mathsf{T}}$
composed of Gaussian white noises with $\langle\hat{\xi}_{n}(t)\rangle=0$
and two-time correlators $\langle\hat{\xi}_{m}(t)\hat{\xi}_{n}(t')\rangle=\mathcal{G}_{mn}\delta(t-t')$,
which define a noise-correlation matrix $\mathcal{G}$. Note that
the number $N$ of independent noises needs not equal $D$ (e.g. in
an optical cavity there are several input vacua per mode, even if
in many instances one can ignore all but one).

Let us remark that the stochastic Langevin equations naturally obtained
from the Schr\"odinger picture through phase-space representations such
as the positive $P$ \citep{PositiveP1,PositiveP2} have the same
form as Eq. (\ref{H-L}), but replacing operators by suitable stochastic
variables. Hence, the theory that we are going to put forward applies
also to such an alternative, but common approach to open quantum systems.

Note that throughout this work we use bold fonts for vectors, e.g.
$\boldsymbol{r}$, which by default correspond to columns with components
denoted by $r_{m}$, so that $\boldsymbol{r}^{\mathsf{T}}$ corresponds to
a row vector; also, a dagger will denote the conjugate-transpose as usual, e.g. $\boldsymbol{r}^{\dagger}\coloneqq\boldsymbol{r}^{\ast\mathsf{T}}$. On the other hand, we use calligraphic fonts for matrices,
e.g. $\mathcal{G}$, whose components we denote by $\mathcal{G}_{mn}$.

We follow the standard linearization procedure that starts by splitting
each operator $\hat{r}_{m}$ as its mean field $\langle\hat{r}_{m}\rangle$
plus a fluctuation $\hat{x}_{m}$, i.e. $\hat{\boldsymbol{r}}=\langle\hat{\boldsymbol{r}}\rangle+\hat{\boldsymbol{x}}$.
In the semiclassical approximation that is commonly adopted, the mean
field $\langle{\hat{\boldsymbol{r}}}\rangle$, to be denoted as $\boldsymbol{r}$,
is ruled by the dynamical system of equations $d\boldsymbol{r}/dt=\textbf{A}\left(\lambda;\boldsymbol{r}\right)$,
obtained from Eq. (\ref{H-L}) by substituting operators by their
mean values and ignoring noises. These correspond to the classical
limit, and we are here interested in the case where such classical
dynamics is periodic, i.e. $\boldsymbol{r}=\boldsymbol{R}(t)$, with
$\boldsymbol{R}\left(t+T\right)=\boldsymbol{R}(t)$ some periodic
function with period $T$. This can happen either when the control
parameter $\lambda$ is periodically modulated in time, or following a dynamical (typically Hopf) bifurcation occurring at some critical value  $\lambda=\lambda_{\textrm{osc}}$
which marks the onset of self-sustained oscillations (see \citep{LC}
for a detailed example).

The dynamics of the fluctuations is governed by the original quantum
Langevin equations (\ref{H-L}), which after linearization with respect
to fluctuations and noises are written as 
\begin{equation}
\frac{d\hat{\boldsymbol{x}}}{dt}=\mathcal{L}(t)\hat{\boldsymbol{x}}+\mathcal{B}(t)\hat{\boldsymbol{\xi}}(t),\label{modelrvector}
\end{equation}
where we denote $\mathcal{B}[\lambda;\boldsymbol{R}(t)]$ simply by
$\mathcal{B}(t)$, and the $(D\times D)$-matrix $\mathcal{L}$ is
the Jacobian of the classical dynamical equations, with elements $\mathcal{L}_{mn}(t)=\left.\partial A_{m}\left(\lambda;\boldsymbol{r}\right)/\partial r_{n}\right|_{\boldsymbol{r}=\boldsymbol{R}(t)}$.
The Jacobian $\mathcal{L}$ depends on the parameters and on the classical
solution, and thus it is explicitly $T$-periodic, $\mathcal{L}(t+T)=\mathcal{L}(t)$, as is the matrix $\mathcal{B}(t)$.
Hence (\ref{modelrvector}) is a non-autonomous dynamical system of
linear equations, which prevents its analytical solving. However application
of Floquet theory allows us to transform Eq. (\ref{modelrvector})
into a system with a time-independent Jacobian, which is more amenable
to analytical or semi-analytical treatments. Let us review here the
procedure, which we will exploit throughout the rest of the work \cite{RefAFloquet1,RefAFloquet2,RefFloquet3,RefFloquet4}.
We start by defining the \textit{principal fundamental matrix} $\mathcal{F}(t)$
through the initial-value problem 
\begin{equation}
\frac{d\mathcal{F}}{dt}=\mathcal{L}(t)\mathcal{F},\quad\mathcal{F}(0)=\mathcal{I}_{D\times D},\label{Fdef}
\end{equation}
where $\mathcal{I}_{D\times D}$ is the $(D\times D)$ identity matrix.
\btxt{Note that we choose the initial time as 0 without loss of generality, because any other choice, e.g. $\mathcal{F}(t_0)=\mathcal{I}_{D\times D}$ with $t_0\neq 0$, is connected to Eqs. (\ref{modelrvector}) and (\ref{Fdef}) by the change of variables $\tilde{t}=t-t_0$ and $\tilde{\boldsymbol{x}}(\tilde{t})=\hat{\boldsymbol{x}}(\tilde{t}+t_0)$, leading to a Floquet problem with associated principal fundamental matrix $\tilde{\mathcal{F}}(\tilde{t})=\mathcal{F}(\tilde{t}+t_0)$}. Next, we construct
a \textit{constant matrix} $\mathcal{M}$ through 
\begin{equation}
e^{\mathcal{M}T}=\mathcal{F}\left(T\right),\label{Mdef}
\end{equation}
which serves to decompose the fundamental matrix in its so-called
Floquet normal form, 
\begin{equation}
\mathcal{F}(t)=\mathcal{P}(t)e^{\mathcal{M}t},\label{Pdef}
\end{equation}
where $\mathcal{P}(t)$ is a $T$-periodic invertible matrix. Defining
a transformed fluctuation vector 
\begin{equation}
\hat{\boldsymbol{s}}(t)\coloneqq\mathcal{P}^{-1}(t)\hat{\boldsymbol{x}}(t)\label{def_s}
\end{equation}
\btxt{the non-autonomous equation (\ref{modelrvector}) with time-periodic coefficients turns into} 
\begin{equation}
\frac{d\hat{\boldsymbol{s}}}{dt}=\mathcal{M}\hat{\boldsymbol{s}}+\mathcal{P}^{-1}(t)\mathcal{B}(t)\hat{\boldsymbol{\xi}}(t),\label{dsdt}
\end{equation}
\btxt{which is an equation with time-independent coefficients and time-dependent forcing}. This constitutes an example of Floquet's theorem.

The system of equations (\ref{dsdt}) can be formally solved in terms
of the eigensystem of matrix $\mathcal{M}$. Let us denote by $\mathcal{S}$
the $(D\times D)$ matrix that diagonalizes $\mathcal{M}$ through
the similarity transformation 
\begin{equation}
\mathcal{S}^{-1}\mathcal{M}\mathcal{S}=\mathcal{D},\qquad\text{with }\mathcal{D}=\left(\begin{array}{ccc}
\mu_{1}\\
 & \ddots\\
 &  & \mu_{D}
\end{array}\right).\label{EigenM}
\end{equation}
The eigenvalues $\lbrace\mu_{\alpha}\rbrace_{\alpha=1}^{D}$ are known
as Floquet (or characteristic) exponents. Note that in previous works
\citep{LC} we have used a slightly less compact notation, where we
defined the set of right and left eigenvectors of $\mathcal{M}$,
satisfying $\mathcal{M}\boldsymbol{v}_{\alpha}=\mu_{\alpha}\boldsymbol{v}_{\alpha}$,
$\boldsymbol{w}_{\alpha}^{\dagger}\mathcal{M}=\mu_{\alpha}\boldsymbol{w}_{\alpha}^{\dagger}$,
and orthonormality relations $\boldsymbol{w}_{\alpha}^{\dagger}\boldsymbol{v}_{\beta}=\delta_{\alpha\beta}$.
These two notations are connected by
\begin{equation}
\mathcal{S}=(\boldsymbol{v}_{1}...\boldsymbol{v}_{D})\quad\text{and}\quad\mathcal{S}^{-1}=\left(\begin{array}{c}
\boldsymbol{w}_{1}^{\dagger}\\
\vdots\\
\boldsymbol{w}_{D}^{\dagger}
\end{array}\right).
\end{equation}
It proves convenient to define the auxiliary matrix
\begin{equation}
\mathcal{K}(t)\coloneqq\mathcal{P}(t)\mathcal{S}.\label{KQmatrices}
\end{equation}
 Upon multiplying (\ref{dsdt}) by $\mathcal{S}^{-1}$ from the left,
and defining the projections \begin{subequations}\label{projections}
\begin{align}
\hat{\boldsymbol{c}}(t) & \coloneqq\mathcal{S}^{-1}\hat{\boldsymbol{s}}(t)=\mathcal{K}^{-1}(t)\hat{\boldsymbol{x}}(t),\label{def_c}\\
\hat{\boldsymbol{n}}(t) & \coloneqq\mathcal{K}^{-1}(t)\mathcal{B}(t)\hat{\boldsymbol{\xi}}(t),\label{ProjNoises}
\end{align}
\end{subequations}we get 
\begin{align}
\frac{d\hat{\boldsymbol{c}}}{dt}=\mathcal{D}\hat{\boldsymbol{c}}+\hat{\boldsymbol{n}}(t),
\end{align}
which are a set of decoupled linear equations for the components of
$\boldsymbol{c}$, whose formal solution can be put as 
\begin{equation}
\hat{c}_{\alpha}(t)=\int_{-\infty}^{t}dt'e^{\mu_{\alpha}(t-t')}\hat{n}_{\alpha}(t').\label{sol_c}
\end{equation}
Here we assumed that all the eigenvalues $\mu_{\alpha}$ have negative
real part (i.e., the analyzed semiclassical state is linearly stable),
hence the integral (\ref{sol_c}) is bounded.

Expressions (\ref{Fdef}), (\ref{Mdef}), (\ref{Pdef}), (\ref{EigenM}),
and (\ref{sol_c}) constitute the basis of our analysis, as they allow
computing the fluctuation vector
\begin{equation}
\hat{\boldsymbol{x}}(t)=\mathcal{K}(t)\boldsymbol{c}(t),\label{cTOr}
\end{equation}
in terms of the noise integrals that depend only on the auxiliary
matrix $\mathcal{K}(t)$ and the Floquet exponents $\lbrace\mu_{\alpha}\rbrace_{\alpha=1}^{D}$.

Note that computing matrix $\mathcal{M}$
is not required at any step. Instead, we can use the so-called monodromy
matrix $\mathcal{F}(T)$, which is diagonalized by the same similarity
transformation (\ref{EigenM}), and possesses eigenvalues $\lbrace\phi_{\alpha}\rbrace_{\alpha=1}^{D}$
related to the Floquet exponents by $\mu_{\alpha}T=\ln{\phi_{\alpha}}$.

\section{Computation of two-time correlations}

Our goal is the computation of physical quantities related to the
quantum fluctuations of the system $\boldsymbol{x}$ around the stable,
periodic semiclassical solution $\boldsymbol{r}=\boldsymbol{R}(t)$.
Within the linearized approximation that we are using, which is equivalent
to assuming the state to be Gaussian \cite{CNB14-Linear,CNBbook}, the most general
quantities that one can consider are two-time correlations, since
for Gaussian distributions any higher-order correlation can be reduced
to products of two-time ones. Hence, the most general correlators
we want to compute are 
\begin{equation}
\mathcal{X}(t,t')\coloneqq\langle\hat{\boldsymbol{x}}(t)\hat{\boldsymbol{x}}^{\mathsf{T}}(t')\rangle,
\end{equation}
where we remind that $\hat{\boldsymbol{x}}$ is a column vector, so
$\mathcal{X}$ is a matrix. As a first result of this work, we provide
here a simple expression for this two-time correlation matrix that
exploits the periodic nature of the problem. We start by using (\ref{cTOr})
to rewrite it as 
\begin{equation}
\mathcal{X}(t,t')=\mathcal{K}(t)\mathcal{C}(t,t')\mathcal{K}^{\mathsf{T}}(t'),\label{Cij}
\end{equation}
where
\begin{align}
\mathcal{C}(t,t')\coloneqq\langle\hat{\boldsymbol{c}}(t)\hat{\boldsymbol{c}}^{\mathsf{T}}(t')\rangle,\label{def_Cab}
\end{align}
are elementary correlations that we work out in Appendix \ref{AppendixA:TwoTimeCorrs}.
We relegate the technical derivations to that appendix, and summarize
here only the final compact expressions. Note first that the
projected noises (\ref{ProjNoises}) are delta correlated as 
\begin{align}
\langle\hat{\boldsymbol{n}}(t)\hat{\boldsymbol{n}}^{\mathsf{T}}(t')\rangle=\mathcal{N}(t')\delta(t-t'),\label{Ndab}
\end{align}
with a projected-noise correlation matrix $\mathcal{N}(t)=\mathcal{K}^{-1}(t)\mathcal{B}(t)\mathcal{G}\mathcal{B}^{\mathsf{T}}(t)\mathcal{K}^{-1\mathsf{T}}(t)$
that is obviously $T$-periodic. With this definition at hand, we
show in Appendix \ref{AppendixA:TwoTimeCorrs} that the correlation
matrix (\ref{def_Cab}) can be worked out to yield the components
\begin{align}
\mathcal{C}_{\alpha\beta}(t,t')=\Upsilon(\mu_{\alpha}+\mu_{\beta})\overline{C}_{\alpha\beta}(t,t'),\label{Cab}
\end{align}
where \begin{subequations}
\begin{align}
\Upsilon(x) & \coloneqq\frac{e^{xT}}{1-e^{xT}},\label{Yota}\\
\overline{C}_{\alpha\beta}(t,t') & \coloneqq\begin{cases}
\Gamma_{\alpha\beta}(t'\bmod{T})e^{\mu_{\alpha}(t-t')}, & t'\le t,\\
\Gamma_{\alpha\beta}(t\bmod{T})e^{\mu_{\beta}(t'-t)}, & t\le t',
\end{cases}\label{overCab}
\end{align}
\end{subequations}with \begin{subequations}\label{Nab} 
\begin{align}
\Gamma_{\alpha\beta}(\tau) & \coloneqq e^{(\mu_{\alpha}+\mu_{\beta})\tau}\left[\nu_{\alpha\beta}(T)+\frac{\nu_{\alpha\beta}(\tau)}{\Upsilon(\mu_{\alpha}+\mu_{\beta})}\right],\\
\nu_{\alpha\beta}(\tau) & \coloneqq\int_{0}^{\tau}dt_{1}\,e^{-(\mu_{\alpha}+\mu_{\beta})t_{1}}\mathcal{N}_{\alpha\beta}(t_{1}).\label{nu}
\end{align}
\end{subequations}

Eqs. (\ref{Cab})-(\ref{Nab}) are the first main result of this
work as they allow us to compute any two-time correlation in terms
of integrals of functions evaluated just in the interval $t\in\left[0,T\right]$.
The importance of this result emerges when long measurement times
are involved, as those required for the computation of spectral densities
(see next section), because, apart from being numerically demanding,
the errors accumulated in finding the fundamental matrix $\mathcal{F}(t)$
at long times can be large enough to invalidate the results.

Note that, for numerical purposes, it is typically more efficient
to evaluate $\nu_{\alpha\beta}(\tau)$ from the equivalent initial-value
problem
\begin{equation}
\dot{\nu}_{\alpha\beta}=e^{-(\mu_{\alpha}+\mu_{\beta})t}\mathcal{N}_{\alpha\beta}(t),\quad\nu_{\alpha\beta}(0)=0,\label{nuDIFF}
\end{equation}
rather than from the integral (\ref{nu}).

\section{Computation of spectral densities}

Another important tool for characterizing quantum fluctuations are
the spectral densities associated to two-time correlations. A relevant
example is the light squeezing spectrum, which is the spectral variance
of the (quantum) noise carried by a light beam, and can be measured
experimentally via balanced homodyne detection \citep{Gea,CNB-PhD} or
alternative correlation measurements \citep{Vogel}. In the usual stationary case, i.e. when two-time
correlations are a function only of the two-time difference, these
densities are just plain Fourier transforms. However, when such correlations
are not stationary one has to use a different definition in order
to match the experimentally detected spectral density \citep{Gea,CNB-PhD},
namely 
\begin{equation}
S(\omega)\coloneqq\frac{1}{T_{d}}\int_{0}^{T_{d}}dt\int_{0}^{T_{d}}dt'O(t,t')e^{i\omega(t-t')},\label{def_S}
\end{equation}
where $O(t,t')$ is the considered two-time correlation and $T_{d}$
is the detection time. In general the measurable densities will be
linear combinations of $S(\omega)$ and $S(-\omega)$, as we will
see later through a practical example. We then consider spectral densities
of the form 
\begin{align}\label{def_Sab}
\mathcal{S}_{\alpha\beta}(\omega;P_{\alpha},P_{\beta})\hspace{-0.5mm}&\coloneqq\frac{1}{T_{d}}\int_{0}^{T_{d}}dt\int_{0}^{T_{d}}dt'P_{\alpha}(t)P_{\beta}(t')
\\
&\hspace{2.5cm}\times\mathcal{C}_{\alpha\beta}(t,t')e^{i\omega(t-t')},\nonumber
\end{align}
obtained upon setting $O(t,t')=P_{\alpha}(t)P_{\beta}(t')\mathcal{C}_{\alpha\beta}(t,t')$
in (\ref{def_S}), being $P_{\alpha}(t)$ and $P_{\beta}(t')$ generic
$T$-periodic functions whose meaning is as follows. When such functions
are chosen as $\mathcal{K}_{m\alpha}(t)$ and $\mathcal{K}_{n\beta}(t')$,
respectively, and summing over $\alpha$ and $\beta$, one can compute
spectral densities corresponding to the correlations $\mathcal{X}_{mn}(t,t')$,
see (\ref{Cij}). The choice $P_{\alpha}(t)=P_{\beta}(t')=1$ is also
interesting as it provides the spectral densities corresponding to
the elementary correlations $\mathcal{C}_{\alpha\beta}(t,t')$, which
in some cases are proportional to measurable quadratures \citep{CNB-PhD,CNB08,CNB09,CNB10,SPOPOrot}.
Finally, when $\hat{\boldsymbol{r}}$ is formed of annihilation and creation operators, with the choice $P_{\alpha}(t)=\Lambda_{\alpha}(t)\mathcal{K}_{m\alpha}(t)$
and $P_{\beta}(t')=\Lambda_{\beta}(t')\mathcal{K}_{n\beta}(t')$,
(\ref{def_Sab}) allows the computation of spectral densities corresponding
to homodyne-detection experiments when the local oscillator is a $T$-periodic
function \cite{SPOPO}, in which case $\Lambda_{\alpha}(t)$ and
$\Lambda_{\beta}(t')$ are proportional to the amplitude (or its complex
conjugate) of that local oscillator.

At first sight it seems easy to solve the problem once the correlation
functions $\mathcal{C}_{\alpha\beta}(t,t')$ have been expressed in
Eqs. (\ref{Cab}) in terms of the first period. However, for long
measurement times, as realistically needed, the integrals are still
numerically demanding and can carry important numerical errors. In
order to avoid this, we have worked out Eq. (\ref{def_Sab}) by exploiting
the properties of the integral's kernel, and managed to simplify it
into a few integrals defined only over a single period. As we did
in the previous section, we relegate the technical derivations to
Appendix \ref{Appendix:SpectralDensities}, presenting here the final
result. Moreover, we focus on the common situation of a long detection
time that contains very many periods, that is, $T_{d}\gg T$. In this
limit, as proven in Appendix \ref{Appendix:SpectralDensities}, the
general spectral density (\ref{def_Sab}) is simplified as
\begin{align}\label{Sab_simplified}
\mathcal{S}_{\alpha\beta}(\omega;P_{\alpha},&P_{\beta})=\frac{\Upsilon(\mu_{\alpha}+\mu_{\beta})}{T}\big[I_{\alpha\beta}^{\text{\large\ensuremath{\lrcorner}}}(\omega)+I_{\alpha\beta}^{\,\rotatebox[origin=c]{180}{\text{\large\ensuremath{\lrcorner}}}}(\omega)
\\
&+\Upsilon(\mu_{\alpha}+i\omega)I_{\alpha\beta}^{\searrow}(\omega)+\Upsilon(\mu_{\beta}-i\omega)I_{\alpha\beta}^{\nwarrow}(\omega)\big],\nonumber
\end{align}
where\begin{subequations}
\begin{align}
I_{\alpha\beta}^{\searrow}(\omega) & \coloneqq\int_{0}^{T}dtP_{\alpha}(t)e^{(\mu_{\alpha}+i\omega)t}
\\
&\hspace{1cm}\times\int_{0}^{T}dt'P_{\beta}(t')\Gamma_{\alpha\beta}(t')e^{-(\mu_{\alpha}+i\omega)t'},\nonumber
\\
I_{\alpha\beta}^{\nwarrow}(\omega) & \coloneqq\int_{0}^{T}dt'P_{\beta}(t')e^{(\mu_{\beta}-i\omega)t'}
\\
&\hspace{1cm}\times\int_{0}^{T}dtP_{\alpha}(t)\Gamma_{\alpha\beta}(t)e^{-(\mu_{\beta}-i\omega)t},\nonumber
\\
I_{\alpha\beta}^{\text{\large\ensuremath{\lrcorner}}}(\omega) & \coloneqq\int_{0}^{T}dt'P_{\beta}(t')\Gamma_{\alpha\beta}(t')e^{-(\mu_{\alpha}+i\omega)t'}
\\
&\hspace{1cm}\times\int_{t'}^{T}dtP_{\alpha}(t)e^{(\mu_{\alpha}+i\omega)t},\nonumber
\\
I_{\alpha\beta}^{\,\rotatebox[origin=c]{180}{\text{\large\ensuremath{\lrcorner}}}}(\omega) & \coloneqq\int_{0}^{T}dtP_{\alpha}(t)\Gamma_{\alpha\beta}(t)e^{-(\mu_{\beta}-i\omega)t}
\\
&\hspace{1cm}\times\int_{t}^{T}dt'P_{\beta}(t')e^{(\mu_{\beta}-i\omega)t'}.\nonumber
\end{align}
\end{subequations}
\btxt{Let us remark that the superindex labeling each integral is not arbitrary, but connected to the original integration domain from where they emerge in the $(t,t')$ space. In particular, in Appendix \ref{Appendix:SpectralDensities} we show that dividing the $(t,t')$ space into a sort of chessboard with squared integration domains of area $T\times T$, $I_{\alpha\beta}^{\searrow}(\omega)$ is the integral to which we can relate all the integrals defined on squares below the $t=t'$ diagonal, hence the `$\searrow$' label.}

Remarkably, again we have been able to write spectral
densities in terms of first-period objects only, which comes with
all the numerical benefits that we highlighted above. Hence, this
is the second main result of our work, which provides a compact way
of evaluating arbitrary spectral densities in periodic systems from
knowledge of the Floquet eigensystem over a single period.

Similarly to what we did in the previous section with $\nu_{\alpha\beta}(t)$
in Eq. (\ref{nuDIFF}), it is useful for numerical efficiency to find
the integrals defined above from their equivalent differential equations.
In the case of $I_{\alpha\beta}^{\searrow}(\omega)$ and $I_{\alpha\beta}^{\nwarrow}(\omega)$,
both are of the integral form $I=\int_{0}^{T}dtf(t)\int_{0}^{T}dt'h(t')$.
Hence, defining two independent initial-value problems\begin{subequations}
\begin{align}
\dot{F} & =f(t),\quad F(0)=0,\\
\dot{H} & =h(t),\quad H(0)=0,
\end{align}
\end{subequations}we get $I=F(T)H(T)$. On the other hand, $I_{\alpha\beta}^{\text{\large\ensuremath{\lrcorner}}}(\omega)$
and $I_{\alpha\beta}^{\,\rotatebox[origin=c]{180}{\text{\large\ensuremath{\lrcorner}}}}(\omega)$
are of the nested type $I=\int_{0}^{T}dtf(t)\int_{t}^{T}dt'h(t')$,
which makes their differential form a bit more intricate, but equally
efficient from a numerical standpoint. In this case, we first solve
the initial-value problem
\begin{equation}
\dot{H}=-h(t),\quad H(T)=0,
\end{equation}
backwards in time in the domain $t\in[0,T]$, and next the initial-value
problem
\begin{equation}
\dot{F}=f(t)H(t),\quad F(0)=0,
\end{equation}
so that $I=F(T)$.

\section{Cross-correlations and cross-spectra with the noise}

In the previous sections we focused on the two-time correlations and
spectral densities of the variables $\hat{\boldsymbol{x}}$ (or, equivalently,
the projections $\hat{\boldsymbol{c}}$). However, in many situations
one also needs objects related to the cross-correlations between the
variables and the noises $\hat{\boldsymbol{\xi}}$. A most prominent
case is related to the evaluation of quantities related to the field
leaking out of the open system by using input-output relations. We
will showcase this in the practical example that we consider in the
next section. This section is then devoted to provide compact expressions
for these type of cross-correlations and spectral densities. Again
we make all technical derivations in Appendix \ref{Appendix_CrossCorrNoise},
and offer here just the final results.

We start by providing the two-time cross-correlators between the projections
$\hat{\boldsymbol{c}}$ and the noises $\hat{\boldsymbol{\xi}}$,
which are easily worked out as\begin{subequations}\label{def_Ccross}
\begin{equation}
\mathcal{C}_{\alpha\beta}^{(c\xi)}(t,t'):=\langle\hat{c}_{\alpha}(t)\hat{\xi}_{\beta}(t')\rangle=\begin{cases}
e^{\mu_{\alpha}(t-t')}\chi_{\alpha\beta}^{(c\xi)}(t'), & t'\leq t
\\
0, & t<t'
\end{cases},
\end{equation}
and
\begin{equation}
\mathcal{C}_{\alpha\beta}^{(\xi c)}(t,t'):=\langle\hat{\xi}_{\alpha}(t)\hat{c}_{\beta}(t')\rangle=\begin{cases}

0, & t'<t
\\
e^{\mu_{\beta}(t'-t)}\chi_{\alpha\beta}^{(\xi c)}(t), & t\leq t'
\end{cases},
\end{equation}
\end{subequations}where we have defined the matrices $\chi^{(c\xi)}(t)=\mathcal{K}^{-1}(t)\mathcal{B}(t)\mathcal{G}$
and $\chi^{(\xi c)}(t)=\mathcal{G}\mathcal{B}^{\mathsf{T}}(t)\mathcal{K}^{-1\mathsf{T}}(t)$. \btxt{Let us remind that $\mathcal{K}$ is the auxiliary matrix defined in Eq. (\ref{KQmatrices}), $\mathcal{B}$ is the matrix multiplying the noise vector $\boldsymbol{\xi}(t)$ in the linearized equations (\ref{modelrvector}), and $\mathcal{G}$ is the matrix defined after Eq. (\ref{H-L}) summarizes the two-time correlators of the noise as $\langle\hat{\boldsymbol{\xi}}(t)\hat{\boldsymbol{\xi}}^\mathsf{T}(t')\rangle=\mathcal{G}\delta(t-t')$.}

The corresponding spectral densities, defined, respectively, by replacing
$\mathcal{C}_{\alpha\beta}(t,t')$ in (\ref{def_Sab}) by $\mathcal{C}_{\alpha\beta}^{(c\xi)}(t,t')$
and $\mathcal{C}_{\alpha\beta}^{(\xi c)}(t,t')$, are worked out in
Appendix \ref{Appendix_CrossCorrNoise}, and take the final form\begin{subequations}\label{def_Scross}
\begin{align}
\mathcal{S}_{\alpha\beta}^{(c\xi)}(\omega;P_{\alpha},P_{\beta}) & =\frac{1}{T}\left[J_{\alpha\beta}^{\text{\large\ensuremath{\lrcorner}}}(\omega)+\Upsilon(\mu_{\alpha}+i\omega)J_{\alpha\beta}^{\searrow}(\omega)\right],\\
\mathcal{S}_{\alpha\beta}^{(\xi c)}(\omega;P_{\alpha},P_{\beta}) & =\frac{1}{T}\left[J_{\alpha\beta}^{\,\rotatebox[origin=c]{180}{\text{\large\ensuremath{\lrcorner}}}}(\omega)+\Upsilon(\mu_{\beta}-i\omega)J_{\alpha\beta}^{\nwarrow}(\omega)\right],
\end{align}
\end{subequations}
with
\begin{subequations}
\begin{align}
J_{\alpha\beta}^{\searrow}(\omega) & \coloneqq\int_{0}^{T}dtP_{\alpha}(t)e^{(\mu_{\alpha}+i\omega)t}
\\
&\hspace{1cm}\times\int_{0}^{T}dt'P_{\beta}(t')\chi_{\alpha\beta}^{(c\xi)}(t')e^{-(\mu_{\alpha}+i\omega)t'},\nonumber
\\
J_{\alpha\beta}^{\nwarrow}(\omega) & \coloneqq\int_{0}^{T}dt'P_{\beta}(t')e^{(\mu_{\beta}-i\omega)t'}
\\
&\hspace{1cm}\times\int_{0}^{T}dtP_{\alpha}(t)\chi_{\alpha\beta}^{(\xi c)}(t)e^{-(\mu_{\beta}-i\omega)t},\nonumber
\\
J_{\alpha\beta}^{\text{\large\ensuremath{\lrcorner}}}(\omega) & \coloneqq\int_{0}^{T}dt'P_{\beta}(t')\chi_{\alpha\beta}^{(c\xi)}(t')e^{-(\mu_{\alpha}+i\omega)t'}
\\
&\hspace{1cm}\times\int_{t'}^{T}dtP_{\alpha}(t)e^{(\mu_{\alpha}+i\omega)t},\nonumber
\\
J_{\alpha\beta}^{\,\rotatebox[origin=c]{180}{\text{\large\ensuremath{\lrcorner}}}}(\omega) & \coloneqq\int_{0}^{T}dtP_{\alpha}(t)\chi_{\alpha\beta}^{(\xi c)}(t)e^{-(\mu_{\beta}-i\omega)t}
\\
&\hspace{1cm}\times\int_{t}^{T}dt'P_{\beta}(t')e^{(\mu_{\beta}-i\omega)t'}.\nonumber
\end{align}
\end{subequations}

\section{Application: Degenerate parametric oscillation beyond the rotating-wave approximation}

As an application of the method developed above, we consider now the
degenerate parametric oscillator as an example. In essence, it consists
of a lossy quantum-mechanical harmonic oscillator whose frequency
is modulated periodically at twice its natural frequency \btxt{(parametrically-driven oscillator)}. This model
serves as the canonical one for the study of quantum squeezing, and
has been traditionally explored experimentally with nonlinear optical
cavities \cite{CNB-PhD}. Since in this context accessible modulation amplitudes are
much smaller than optical frequencies, one can perform a rotating-wave
approximation that maps the problem to an effective time-independent
one. In contrast, modern implementations based on low-frequency oscillators
(e.g., in superconducting circuits \cite{Legtas15} or optomechanical devices \cite{CNB-DPO-OM}) allow to explore the regime where the modulation amplitudes are a significant
fraction of the oscillation frequencies. Under such conditions, the
predictions derived within the rotating-wave approximation require
corrections, and it is our purpose to study these here.

\subsection{The model}\label{DPOmodel}

Consider an oscillator of mass $m$ and intrinsic frequency $\Omega$,
with position $\hat{q}$ and momentum $\hat{p}$, such that $[\hat{q},\hat{p}]=i\hbar$.
We can describe the modulated case by the Hamiltonian 
\begin{equation}
\hat{H}(t)=\frac{\hat{p}^{2}}{2m}+\frac{m\Omega^{2}}{2}\left[1+\varepsilon\sin(2\Omega t)\right]\hat{q}^{2},\label{Hqp}
\end{equation}
with (normalized) modulation amplitude $\varepsilon$. Let us write
the position and momentum in terms of annihilation and creation operators
as 
\begin{equation}
\hat{q}=\sqrt{\frac{\hbar}{2m\Omega}}(\hat{a}^{\dagger}+\hat{a}),\hspace{5mm}\hat{p}=\sqrt{\frac{\hbar m\Omega}{2}}i(\hat{a}^{\dagger}-\hat{a}),
\end{equation}
with $[\hat{a},\hat{a}^{\dagger}]=1$\btxt{. Let us consider the slowly-varying operator $\tilde{a}(t)\coloneqq e^{i\Omega t}\hat{a}(t)$, where $\hat{a}(t)$ is the Heisenberg-picture operator. This operator evolves according to $i\hbar\partial_t\tilde{a}=[\tilde{a},\tilde{H}(t)]$, with rotating-picture Hamiltonian}
\begin{equation}
\tilde{H}(t) =\frac{\hbar\Omega\varepsilon}{2}\sin(2\Omega t)\btxt{\tilde{a}}^{\dagger}\btxt{\tilde{a}}+\frac{i\hbar\Omega\varepsilon}{8}\left[\left(1-e^{4i\Omega t}\right)\btxt{\tilde{a}}^{\dagger2}\hspace{-1mm}-\hspace{-1mm}\mathrm{H.c.}\right].\label{HDPO}
\end{equation}
In the limit $\varepsilon\ll1$, one
can invoke the rotating-wave approximation, which allows neglecting
the rapidly-oscillating terms, leading to the time-independent Hamiltonian
$\tilde{H}\approx i\hbar\Omega\varepsilon(\btxt{\tilde{a}}^{\dagger2}-\btxt{\tilde{a}}^{2})/8$.
This is the usual Hamiltonian employed to analyze degenerate parametric
oscillators. Here, in contrast, we use the theory developed in the
previous sections to study the full Hamiltonian (\ref{HDPO}).

\begin{figure*}
\includegraphics[width=1\textwidth]{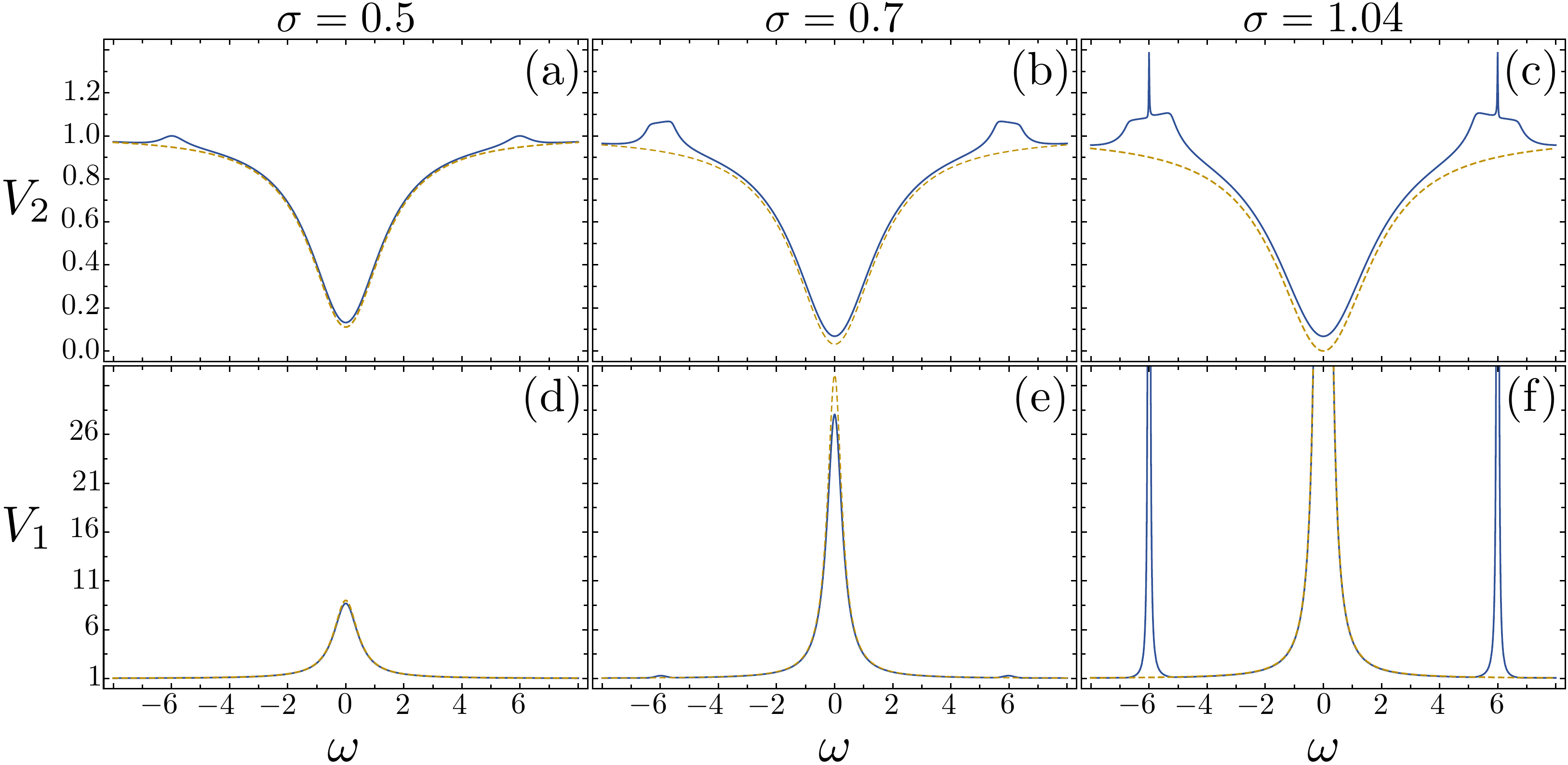}\caption{Eigenvalues $\{V_{1},V_{2}\}$ of the spectral covariance matrix $\mathcal{V}(\omega)$
as a function of the dimensionless detection frequency $\omega$,
for different values of the normalized modulation amplitude $\sigma$.
These eigenvalues determine the squeezing properties of the system,
and can be measured experimentally via homodyne detection. The solid
blue line corresponds to the results provided by our Floquet-based
theory for a quality factor $Q=3$, while the dashed yellow line corresponds
to the standard rotating-wave approximation ($Q\rightarrow\infty$)
for the same value of $\sigma$, except in (c) and (f), where $\sigma=1$
for the latter (it cannot be larger than one, because under the rotating-wave
approximation the system is unstable in such case). Note that $V_{2}<1$
around $\omega=0$, signaling squeezing in the corresponding quadrature.
\label{Fig:Spectra}}
\end{figure*}

In order to include losses, we consider the interaction between the
oscillator and a bosonic environment at zero temperature. Assuming
that the standard Born-Markov approximation holds (see below for further discussion on this point), one can integrate
out the environment leading to the quantum Langevin equation 
\begin{equation}
\frac{d\tilde{a}}{dt}=-\gamma[1+2i\sigma\sin(2\Omega t)]\tilde{a}+\gamma\sigma(1-e^{4i\Omega t})\tilde{a}^{\dagger}+\sqrt{2\gamma}\hat{a}_{\mathrm{in}}(t),\label{QLan}
\end{equation}
where we have defined the normalized modulation amplitude $\sigma\coloneqq\varepsilon\Omega/4\gamma$,
and $\hat{a}_{\mathrm{in}}(t)$ is the so-called input operator, which
is Gaussian and characterized by the following statistical properties:
\begin{equation}
\langle\hat{a}_{\mathrm{in}}(t)\rangle=0=\langle\hat{a}_{\mathrm{in}}^{\dagger}(t)\hat{a}_{\mathrm{in}}(t')\rangle,\hspace{5mm}\langle\hat{a}_{\mathrm{in}}(t)\hat{a}_{\mathrm{in}}^{\dagger}(t')\rangle=\delta(t-t').\label{InputStat}
\end{equation}
Let us also remark that \btxt{the slowly-varying operator $\tilde{a}(t)$} is actually the \btxt{one} that homodyne detection is sensitive to, so \btxt{this is the one} we will use to compute the relevant spectral densities, as explained below.

It is convenient to introduce the dimensionless time $\tilde{t}\coloneqq\gamma t$,
which we adopt in the following but removing the tilde for notational
simplicity. Let us further define the vectors $\hat{\boldsymbol{a}}\coloneqq(\btxt{\tilde{a}},\btxt{\tilde{a}}^\dagger)^{\mathsf{T}}$
and $\hat{\boldsymbol{a}}_{\mathrm{in}}\coloneqq(\hat{a}_{\mathrm{in}},\hat{a}_{\mathrm{in}}^{\dagger})^{\mathsf{T}}$,
from which we build the quadrature vectors $\hat{\boldsymbol{x}}\coloneqq\mathcal{T}\hat{\boldsymbol{a}}$
and $\hat{\boldsymbol{x}}_{\mathrm{in}}\coloneqq\mathcal{T}\hat{\boldsymbol{a}}_{\mathrm{in}}/\sqrt{\gamma}$,
with $\mathcal{T}\coloneqq\tiny{\left(\begin{array}{cc}
1 & 1\\
-i & i
\end{array}\right)}$. In terms of quadratures, Eq. (\ref{QLan}) is then written as the
linear system 
\begin{equation}
\frac{d\hat{\boldsymbol{x}}}{dt}=\mathcal{L}(t)\hat{\boldsymbol{x}}+\sqrt{2}\hat{\boldsymbol{x}}_{\mathrm{in}}(t),\label{LinearFloquetDPO}
\end{equation}
which has the form of a Floquet problem (\ref{modelrvector}), with
the identifications $\mathcal{B}=\sqrt{2}\mathcal{I}_{2\times2}$,
$\hat{\boldsymbol{\xi}}=\hat{\boldsymbol{x}}_{\mathrm{in}}$, $\mathcal{G}=\tiny{\left(\begin{array}{cc}
1 & i\\
-i & 1
\end{array}\right)}$, and Jacobian $\mathcal{L}(t)=\mathcal{L}_\text{RWA}+\mathcal{L}_\text{non-RWA}(t)$ with
\begin{subequations}\label{L_DPO}
\begin{align}
\mathcal{L}_\text{RWA}&\coloneqq\left(\begin{array}{cc}
-1+\sigma & 0\\
0 & -1-\sigma
\end{array}\right),
\\
\mathcal{L}_\text{non-RWA}\hspace{-0.5mm}&\coloneqq\hspace{-0.5mm}\sigma\hspace{-0.5mm}\left(\hspace{-1mm}\begin{array}{cc}
-\cos(4Qt) & 8\cos(Qt)\sin^{3}(Qt)\\
-8\cos^{3}(Qt)\sin(Qt) & \cos(4Qt)
\end{array}\hspace{-1mm}\right)\hspace{-1mm}.
\end{align}
\end{subequations}
This Jacobian has periodicity $T=\pi/Q$ in terms of the normalized frequency
$Q\coloneqq\Omega/\gamma$, which coincides with the resonator quality factor.

Note that we have obtained a linear system of equations directly,
because our initial Hamiltonian (\ref{Hqp}) was quadratic. This is
however an idealization that works only in a limited range of parameters,
whose breakdown is signaled by the equations becoming unstable. For
example, within the common rotating-wave approximation valid when
$\varepsilon=4\sigma/Q\ll1$ as mentioned above, and obtained from
(\ref{LinearFloquetDPO}) by neglecting the oscillatory terms $\mathcal{L}_{\text{non-RWA}}(t)$
in (\ref{L_DPO}), the Jacobian takes the diagonal form $\mathcal{L}_{\text{RWA}}$
with eigenvalues $-(1\pm\sigma)$. Hence, this idealized linear picture
is valid only for $\sigma<1$. Beyond such point, the modulation cannot
be treated as a given $\varepsilon\sin(2\Omega t)$ term anymore,
and needs a dynamical treatment of its own, for example as a dynamical
variable that feels some backaction from the oscillator (known as
pump depletion in optical implementations). Similar behavior is to
be expected beyond the rotating-wave approximation, but this time
signaled by the real part of some Floquet exponent $\mu_{\alpha}$ becoming positive.

Finally, let us comment on the validity of Eq. (\ref{QLan}) as a model for the effect of the environment onto the oscillator. Technically, this simple quantum Langevin equation is bound to break down for sufficiently small $Q$ and large $\varepsilon$, when Born-Markov conditions can no longer be ensured. More refined and complex open models can be derived in these limits \cite{Hanggi0,Hanggi1,Hanggi2}, but in order to illustrate our Floquet-based method, we will stick with the simple model of Eq. (\ref{QLan}), commenting on the effects that it predicts as we depart from the ideal $Q\gg 1$ and $\varepsilon\ll 1$ conditions traditionally considered in the literature.

\subsection{Spectral covariance matrix}

In order to understand the squeezing properties of this system, we
will consider the spectral covariance matrix, which is the standard
object recovered via homodyne detection of the excitations that leak
out of the oscillator (e.g., the light exiting the cavity through
a partially transmissive mirror in a degenerate parametric oscillator).
Introducing the output operator 
\begin{equation}
\hat{\boldsymbol{x}}_{\mathrm{out}}(t)=\sqrt{2}\hat{\boldsymbol{x}}(t)-\hat{\boldsymbol{x}}_{\mathrm{in}}(t),\label{in/out}
\end{equation}
the spectral covariance matrix is defined as 
\begin{equation}
\mathcal{V}(\omega)\coloneqq\frac{1}{4}\left[\mathcal{A}(\omega)+\mathcal{A}(-\omega)+\mathcal{A}^{\mathsf{T}}(\omega)+\mathcal{A}^{\mathsf{T}}(-\omega)\right],\label{Vx}
\end{equation}
with 
\begin{equation}
\mathcal{A}(\omega)\coloneqq\lim_{T_{d}\rightarrow\infty}\frac{1}{T_{d}}\int_{0}^{T_{d}}\hspace{-1.5mm}dt\int_{0}^{T_{d}}\hspace{-1.5mm}dt'\langle\hat{\boldsymbol{x}}_{\mathrm{out}}(t)\hat{\boldsymbol{x}}_{\mathrm{out}}^{\mathsf{T}}(t')\rangle e^{i\omega(t-t')}.\label{Aw}
\end{equation}
We remind that we are working with a dimensionless time, and therefore,
the detection frequency $\omega$ in this equation is also dimensionless,
with the real detection frequency given by $\gamma\omega$. The spectral
covariance matrix (\ref{Vx}) is subject, for all $\omega$, to the
usual constrains of the standard covariance matrix of Gaussian states
\citep{CNBbook,Patera20}. For example, it is real, symmetric, and must posses
positive eigenvalues (corresponding to the spectral density of the
normal quadratures of the problem), and it satisfies the condition $\det\{\mathcal{V}(\omega)\}\geq 1$ linked to Heisenberg's uncertainty relations.

Note that $\mathcal{A}(\omega)$ has the same form as the generic
spectral density that we defined in (\ref{def_S}), just replacing
the generic correlation function $O(t,t')$ by the output correlation
matrix $\mathcal{C}_{\mathrm{out}}(t,t')\coloneqq\langle\hat{\boldsymbol{x}}_{\mathrm{out}}(t)\hat{\boldsymbol{x}}_{\mathrm{out}}^{\mathsf{T}}(t')\rangle$.
Hence, we now proceed to rewrite it in terms of the spectral densities
that we have defined in the previous sections. First, note that $\mathcal{C}_{\mathrm{out}}(t,t')$
can be written in terms of the previously defined correlations (\ref{def_Cab})
and (\ref{def_Ccross}) as
\begin{align}
\mathcal{C}_{\mathrm{out}}(t,t')=&2\mathcal{K}(t)\mathcal{C}(t,t')\mathcal{K}(t')^{\mathsf{T}}-\sqrt{2}\mathcal{K}(t)\mathcal{C}^{(c\xi)}(t,t')
\\
&-\sqrt{2}\mathcal{C}^{(\xi c)}(t,t')\mathcal{K}(t')^{\mathsf{T}}+\mathcal{G}\delta(t-t'),\nonumber
\end{align}
where we have used (\ref{in/out}), (\ref{cTOr}), and the two-time
correlators of the noises as defined after Eq. (\ref{H-L}). Using
now the definitions for the spectral densities that we introduced
in (\ref{def_Sab}) and (\ref{def_Scross}), the components of $\mathcal{A}(\omega)$
are rewritten as
\begin{align}
\mathcal{A}_{mn}(\omega)&=\mathcal{G}_{mn}+2\sum_{\alpha\beta=1}^{2}\mathcal{S}_{\alpha\beta}(\omega;\mathcal{K}_{m\alpha},\mathcal{K}_{n\beta})
\\
&+\sqrt{2}\sum_{\alpha=1}^{2}\left[\mathcal{S}_{\alpha n}^{(c\xi)}(\omega;\mathcal{K}_{m\alpha},1)+\mathcal{S}_{m\alpha}^{(\xi c)}(\omega;1,\mathcal{K}_{n\alpha})\right],\nonumber
\end{align}
an expression that is readily evaluated using the results of the previous
sections. Specifically, we first solve the Floquet problem (\ref{LinearFloquetDPO})
numerically, that is, we determine the Floquet exponents $\{\mu_{\alpha}\}_{\alpha=1,2}$
and $\mathcal{K}(t)$ over one period, and then use the simplified
expressions of the spectral densities as given in (\ref{Sab_simplified})
and (\ref{def_Scross}).

\subsection{Squeezing properties}

Let us start discussing the results within the rotating-wave approximation.
As mentioned above, in this limit the Jacobian in Eq. (\ref{L_DPO})
is time-independent and has the diagonal form $\mathcal{L}_{\text{RWA}}$.
The particularization of the expressions above to such case easily
leads to the following well-known expression for the spectral covariance
matrix of Eq. (\ref{Vx}):
\begin{equation}
\mathcal{V}^{\text{RWA}}(\omega)=\Bigg(\begin{array}{cc}
V_{1}^{\text{RWA}}(\omega) & 0\\
0 & V_{2}^{\text{RWA}}(\omega)
\end{array}\Bigg),
\end{equation}
with \begin{subequations} 
\begin{align}
V_{1}^{\text{RWA}}(\omega)=1+\frac{4\sigma}{(1-\sigma)^{2}+\omega^{2}},\\
V_{2}^{\text{RWA}}(\omega)=1-\frac{4\sigma}{(1+\sigma)^{2}+\omega^{2}}.
\end{align}
\end{subequations} For $\sigma=0$ this is just the covariance matrix
of vacuum for all $\omega$ as expected, as in the absence of modulation,
the oscillator simply relaxes to its ground state. As $\sigma$ increases,
$V_{1}(\omega)$ gets larger and larger, while $V_{2}(\omega)$ gets
smaller and smaller, corresponding to quantum squeezing in the momentum
quadrature. Eventually, at $\sigma=1$ (the so-called ``threshold''),
we get $V_{2}(\omega=0)=0$ and $V_{1}(\omega=0)=\infty$, signaling
perfect momentum squeezing, and the breakdown of our ideal linear
model. Note that the system remains in a minimum uncertainty state
for all $\sigma$, since $\det\{\mathcal{V}(\omega)\}=V_{1}(\omega)V_{2}(\omega)=1$.

\begin{figure}
\includegraphics[width=0.4\textwidth]{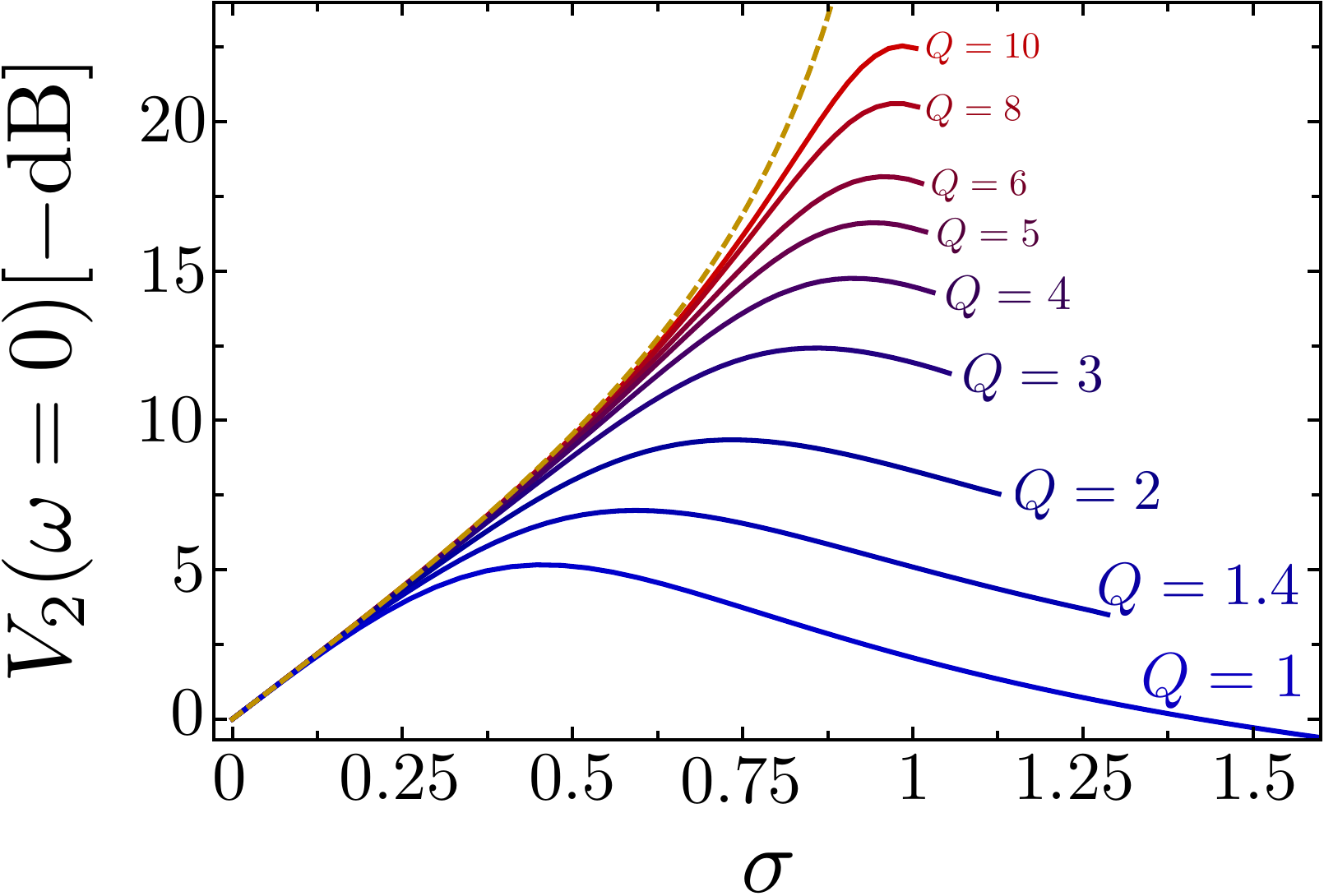}\caption{Zero-frequency spectrum of the squeezed quadrature, $V_{2}(\omega=0)$,
as a function of the normalized modulation amplitude $\sigma$. Different
solid lines correspond to different values of the quality factor $Q$,
with the dashed yellow line showing the rotating-wave approximation
$(Q\rightarrow\infty)$ limit. Note that $V_{2}$ is provided in $-$dB (i.e. we plot $-10\log_{10}V_2$), and hence larger values correspond to better squeezing. Note also
that for finite $Q$ the squeezing is maximum far from the instability
(further the smaller $Q$ is), which for each value of $Q$ corresponds
to the value of $\sigma$ where the curve halts. We show the optimal
squeezing and the corresponding modulation amplitude as functions
of $Q$ in Fig. \ref{Fig:MinV}. \label{Fig:ZeroFreqV}}
\end{figure}

\begin{figure}
\includegraphics[width=0.3\textwidth]{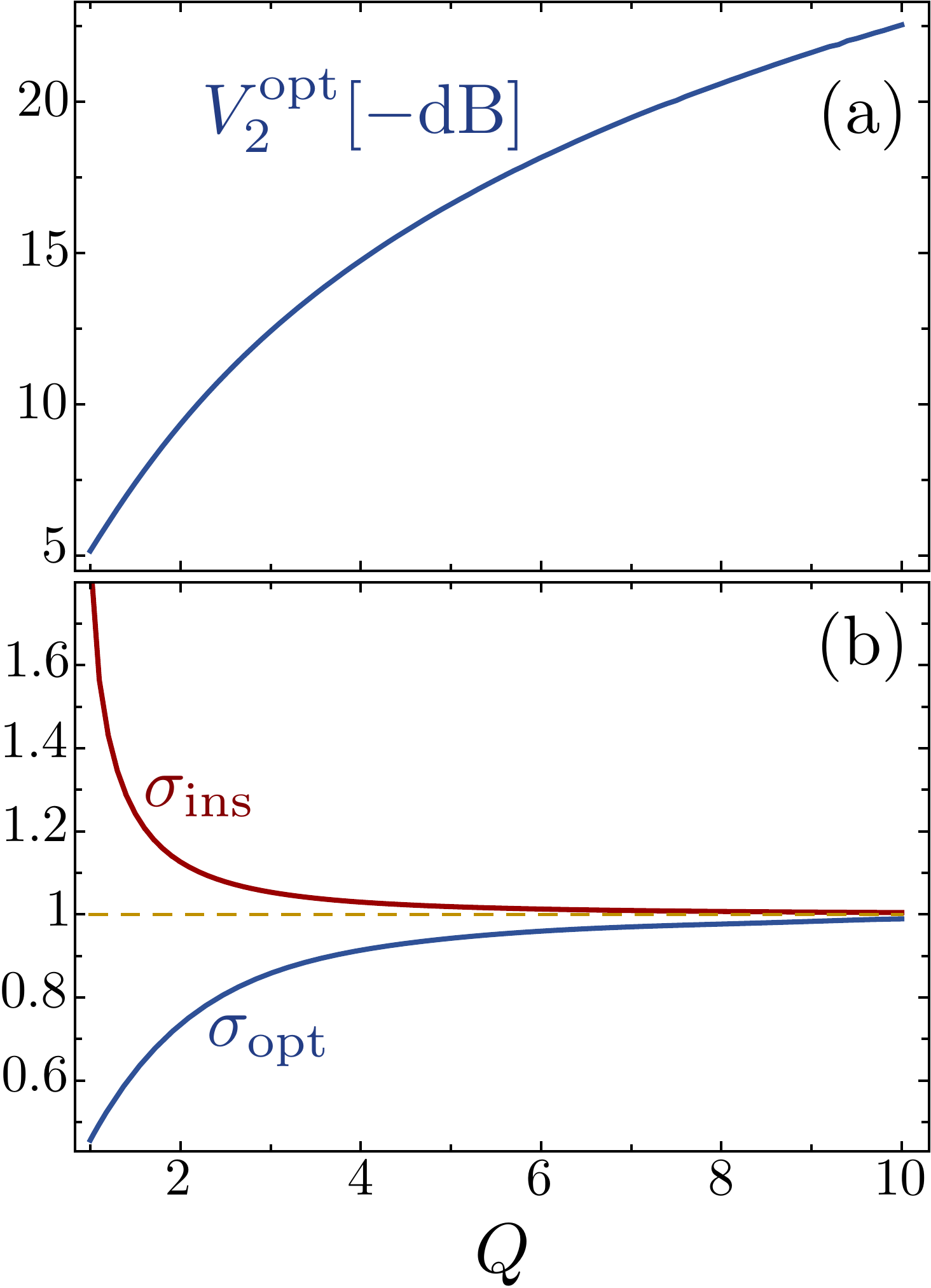}\caption{Optimal squeezing $V_{2}^{\text{opt}}$ (a) and corresponding normalized
modulation amplitude $\sigma_{\text{opt}}$ (b) as a functions of
$Q$. We also show in (b) the value $\sigma_{\text{ins}}$ of the
normalized modulation amplitude for which the system becomes unstable
(the largest Floquet eigenvalue real part vanishes). As in the previous figure, note that $V_2^\text{opt}$ is provided in $-$dB, so larger values correspond to better squeezing \label{Fig:MinV}}
\end{figure}

In this work we have studied the deviations of the full $\mathcal{V}(\omega)$
with respect to this rotating-wave picture. In particular, we summarize
our main results through Figs. \ref{Fig:Spectra} to \ref{Fig:MinV}.
Following the notation introduced above within the rotating-wave approximation,
let us denote by $\{V_{j}(\omega)\}_{j=1,2}$ the eigenvalues of the
spectral covariance matrix $\mathcal{V}(\omega)$ with $V_{1}>V_{2}$
for definiteness. In Fig. \ref{Fig:Spectra} we plot these as a function
of the dimensionless detection frequency $\omega$, for different
values of $\sigma$ (as indicated in the figure) and $Q=3$ (similar
behavior is found for any other value of $Q$). The first
thing that we can appreciate from Figs. \ref{Fig:Spectra}a-c is that
even for a finite $Q$, the optimal squeezing is still found at $\omega=0$,
and is degraded with respect to its rotating-wave value, that is,
$V_{2}(\omega)>V_{2}^{\text{RWA}}(\omega)$. In addition, the spectra
show sidebands at $\omega=\pm2nQ$, with $n\in\mathbb{N}$ (of which we only show $n=\pm 1$ in the plot), as expected for an output field carrying a modulation of period $T=\pi/Q$. The sidebands are relatively broad, and have a shape that departs more and more from Lorentzian as $\sigma$ approaches the instability at which a Floquet eigenvalue becomes zero. We denote such value of $\sigma$ by $\sigma_{\text{ins}}$, which we show in Fig. \ref{Fig:MinV}b as a function of $Q$. Remarkably, once very close to the unstable point, the sidebands of $V_{2}$ develop a secondary sharper peak that diverges at $\sigma=\sigma_{\text{ins}}$ (see Fig. \ref{Fig:Spectra}c).
Let us remark that the sidebands do not show squeezing for any value
of the parameters; on the contrary, they simply add noise. Moreover,
we have also found that the oscillator is not in a minimum uncertainty
state anymore, that is, $V_{1}(\omega)V_{2}(\omega)>1$ for any finite $Q$. Of course,
for any value of the rest of parameters, the product $V_{1}V_{2}$
approaches 1 as $Q$ increases.

Knowing that maximum squeezing occurs at $\omega=0$, in Fig. \ref{Fig:ZeroFreqV}
we plot $V_{2}(\omega=0)$ as a function of $\sigma$ for different
values of $Q$. Note that we plot it in $-$dB units, defined as $-10\text{log}_{10}V_{2}$,
such that higher values correspond to larger squeezing, with 10dB
equivalent to 90\% of quantum noise reduction or $V_{2}=0.1$. Contrary
to the rotating-wave case, squeezing is not maximized at $\sigma=\sigma_{\text{ins}}$,
but at an optimal value $\sigma_{\text{opt}}$ that can be rather
small for small $Q$. This is appreciated in Fig. \ref{Fig:MinV}b,
where we plot $\sigma_{\text{opt}}$ as a function of $Q$, which
of course tends to 1 (the rotating-wave instability) as $Q\rightarrow\infty$. Note also that even
for moderate values of $Q$ the optimal squeezing is quite large,
e.g., $\sim$10dB at $Q=2$, as shown in Fig. \ref{Fig:MinV}a, so
our theory shows that squeezing in parametric oscillation is quite
robust against the quality of the oscillator and the modulation amplitude.

Let us remark, however, that all these predictions rely on the validity of Eq. (\ref{QLan}) as a model for a parametrically-driven oscillator relaxing to its environment. As mentioned at the end of Section \ref{DPOmodel}, this model is expected to break down for sufficiently small $Q$ and large $\varepsilon$, which is precisely where the differences with conventional rotating-wave results become more easily visible. Hence, an interesting question that we will consider in the future is how more refined models \cite{Hanggi0,Hanggi1,Hanggi2} may affect this prediction and how it competes with other effects such as pump depletion, which also limit the squeezing close to the instability.


\section{Conclusions}

In this work we have provided an efficient tool for the evaluation of two-time correlation functions and related spectral densities of time-periodic open quantum systems. In particular, using an approach based on the Floquet theorem, we have shown that these quantities can all be related to simple integrals over a single period, which can be efficiently evaluated. Among other applications, this provides a compact and robust tool for the systematic analysis of the corrections that may arise when generating effective dynamics via periodic modulations. In addition, it is a tool that will find applications in the determination of the quantum properties of systems undergoing limit-cycle motion with the corresponding spontaneous breaking of time-translational invariance.

Let us remark that this method provides an alternative to the direct application of Floquet's theorem at the master equation level \cite{Mintert15,Dai16}, but only for systems that can be linearized. While the latter can certainly be a limitation, it comes with the advantage that no extra conditions are required on the period $T$. In contrast, this period has to be shorter than the time scale of the stroboscopic dynamics in order for the master equation approach to be of practical use, since it typically relies on some kind of perturbative expansion in powers of $T$ \cite{Mintert15,Dai16}. Moreover, in combination with phase-space stochastic Langevin equations, our method is applicable to systems that present self-oscillatory behavior in spite of being described by a master equation with time-independent coefficients \cite{LC}.

As a testbed for the method, we have studied the quantum properties of a damped parametrically-driven oscillator under conditions where the rotating-wave approximation cannot be invoked. This regime is easily attainable nowadays in low-frequency superconducting or mechanical oscillators that work in the quantum regime. Our results show that even for relatively large modulation amplitudes or low-quality oscillators, large levels of squeezing prevail. However, the optimal squeezing levels occur for a modulation amplitude far below the oscillation instability, which is where rotating-wave results predict optimal squeezing. Note that our main goal with this example was to present the Floquet-based method through a characteristic open model that most people working on quantum optics would be familiar with, rather than performing an exhaustive and rigorous analysis of a parametrically-driven oscillator interacting with its environment. In particular, the model we have used is expected to fail for extremely bad oscillators and large modulation amplitudes, for which our predictions will need to be confronted with more suitable models.

\begin{acknowledgments}
This work was funded by the Spanish Ministerio de Ciencia, Innovaci\'{o}n
y Universidades, Agencia Estatal de Investigaci\'{o}n, and the European
Union ``Fondo Europeo de Desarrollo Regional'' (FEDER) through projects
FIS2014-60715-P and FIS2017-89988-P. CNB acknowledges additional
support from a Shanghai talent program and from the Shanghai Municipal
Science and Technology Major Project (Grant No. 2019SHZDZX01).
\end{acknowledgments}

\appendix

\section{Working out two-time correlators\label{AppendixA:TwoTimeCorrs}}

In this appendix we show how to obtain expression (\ref{Cab}) for
the elementary two-time correlators of the projections $\hat{c}_{\alpha}(t)$.
Our starting point is the definition (\ref{def_Cab}), with components
$\mathcal{C_{\alpha\beta}}(t,t')=\langle c_{\alpha}(t)c_{\beta}(t')\rangle$,
which using (\ref{sol_c}), can be expressed as 
\begin{align}
\mathcal{C}_{\alpha\beta}(t,t')=\int_{-\infty}^{t}\hspace{-4mm}dt_{1}\int_{-\infty}^{t'}\hspace{-4mm}dt_{2}e^{\mu_{\alpha}(t-t_{1})+\mu_{\beta}(t'-t_{2})}\langle\hat{n}_{\alpha}(t_{1})\hat{n}_{\beta}(t_{2})\rangle,\label{Cab_1-1}
\end{align}
which is further simplified into 
\begin{equation}
\mathcal{C}_{\alpha\beta}(t,t')=e^{\mu_{\alpha}t+\mu_{\beta}t'}\int_{-\infty}^{\min(t,t')}dt_{1}e^{-(\mu_{\alpha}+\mu_{\beta})t_{1}}\mathcal{N}_{\alpha\beta}(t_{1}),
\end{equation}
where we have used the noise correlators of Eq. (\ref{Nab}) and integrated
out the delta function. Next, we use the periodicity of matrix $\mathcal{N}$,
which suggests writing the above integral as a sum of integrals extended
over consecutive periods, namely 
\begin{align}
\mathcal{C}_{\alpha\beta}(t,t')&=e^{\mu_{\alpha}t+\mu_{\beta}t'}
\\
&\times\sum_{n=0}^{\infty}\int_{\min{(t,t')}-(n+1)T}^{\min{(t,t')}-nT}dt_{1}e^{-(\mu_{\alpha}+\mu_{\beta})t_{1}}\mathcal{N}_{\alpha\beta}(t_{1}).\nonumber
\end{align}
Performing the variable change $t_{2}=t_{1}-\min(t,t')+(n+1)T$, and
using $\mathcal{N}_{\alpha\beta}(t+T)=\mathcal{N}_{\alpha\beta}(t)$,
we obtain 
\begin{align}\label{Cab_aux-1}
\mathcal{C}_{\alpha\beta}(t,t')&=e^{\mu_{\alpha}[t-\min(t,t')]+\mu_{\beta}[t'-\min(t,t')]}
\\
&\hspace{5mm}\times\Upsilon(\mu_{\alpha}+\mu_{\beta})\Gamma_{\alpha\beta}(\tau),\nonumber
\end{align}
where we defined\begin{subequations} 
\begin{align}
\Upsilon(x) & \coloneqq\sum_{n=0}^{\infty}e^{(n+1)xT}=\frac{e^{xT}}{1-e^{xT}},\label{Upsilon-1}\\
\Gamma_{\alpha\beta}(\tau) & \coloneqq\int_{0}^{T}dt_{2}e^{-(\mu_{\alpha}+\mu_{\beta})t_{2}}\mathcal{N}_{\alpha\beta}(t_{2}+\tau),\label{def_Nab-1}\\
\tau & \coloneqq\min(t,t')\bmod T.\label{def_tau-1}
\end{align}
\end{subequations}Expression (\ref{Cab_aux-1}) can be rewritten
as
\begin{align}
\mathcal{C}_{\alpha\beta}(t,t')=\Upsilon(\mu_{\alpha}+\mu_{\beta})\overline{C}_{\alpha\beta}(t,t'),\label{Cab-1}
\end{align}
where 
\begin{align}
\overline{C}_{\alpha\beta}(t,t') & =\begin{cases}
\Gamma_{\alpha\beta}(t'\bmod{T})e^{\mu_{\alpha}(t-t')}, & t'\le t
\\
\Gamma_{\alpha\beta}(t\bmod{T})e^{\mu_{\beta}(t'-t)}, & t\le t'
\end{cases},\label{overCab-1}
\end{align}
which has the same form as Eq. (\ref{Cab}) in the main text, except
for the fact that the integral (\ref{def_Nab-1}) still requires knowledge
of $\mathcal{N}_{\alpha\beta}$ outside the first period because of
its augmented argument. In order to keep the evaluation restricted
to the first period, the integral can be worked out as we explain
next. First, we perform the variable change $t_{3}=t_{2}+\tau$, and
split the resulting integral as 
\begin{align}\label{Fab0}
\Gamma_{\alpha\beta}(\tau)&=e^{(\mu_{\alpha}+\mu_{\beta})\tau}\Bigg[\int_{\tau}^{T}dt_{3}e^{-(\mu_{\alpha}+\mu_{\beta})t_{3}}\mathcal{N}_{\alpha\beta}(t_{3})
\\
&\hspace{1cm}+\int_{T}^{T+\tau}dt_{3}e^{-(\mu_{\alpha}+\mu_{\beta})t_{3}}\mathcal{N}_{\alpha\beta}(t_{3})\Bigg].\nonumber
\end{align}
Next we perform the change of variable $t_{4}=t_{3}-T$ in the second
integral, which is the one that extends beyond the first period. Noting
that $\mathcal{N}_{\alpha\beta}(t_{4}+T)=\mathcal{N}(t_{4})$, we
obtain 
\begin{align}
\Gamma_{\alpha\beta}(\tau)&=e^{(\mu_{\alpha}+\mu_{\beta})\tau}\Bigg[\int_{\tau}^{T}dt_{3}e^{-(\mu_{\alpha}+\mu_{\beta})t_{3}}\mathcal{N}_{\alpha\beta}(t_{3})\\
&\hspace{3mm}+e^{-(\mu_{\alpha}+\mu_{\beta})T}\int_{0}^{\tau}dt_{4}e^{-(\mu_{\alpha}+\mu_{\beta})t_{4}}\mathcal{N}_{\alpha\beta}(t_{4})\Bigg].\nonumber
\end{align}
Finally writing the first integral as $\int_{\tau}^{T}=\int_{0}^{T}-\int_{0}^{\tau}$,
and renaming the dummy variables $t_{3}$ and $t_{4}$ as $t$, we
end up with 
\begin{align}\label{Fab1}
\Gamma_{\alpha\beta}(\tau)&=e^{(\mu_{\alpha}+\mu_{\beta})\tau}\Bigg[\int_{0}^{T}dt\hspace{2pt}e^{-(\mu_{\alpha}+\mu_{\beta})t}\mathcal{N}_{\alpha\beta}(t)
\\
&+\left(e^{-(\mu_{\alpha}+\mu_{\beta})T}-1\right)\int_{0}^{\tau}dt\hspace{2pt}e^{-(\mu_{\alpha}+\mu_{\beta})t}\mathcal{N}_{\alpha\beta}(t)\Bigg],\nonumber
\end{align}
which coincides with Eq. (\ref{Nab}) in the main text.

\section{Working out spectral densities\label{Appendix:SpectralDensities}}

\begin{figure}[t]
\centering \includegraphics[width=\columnwidth]{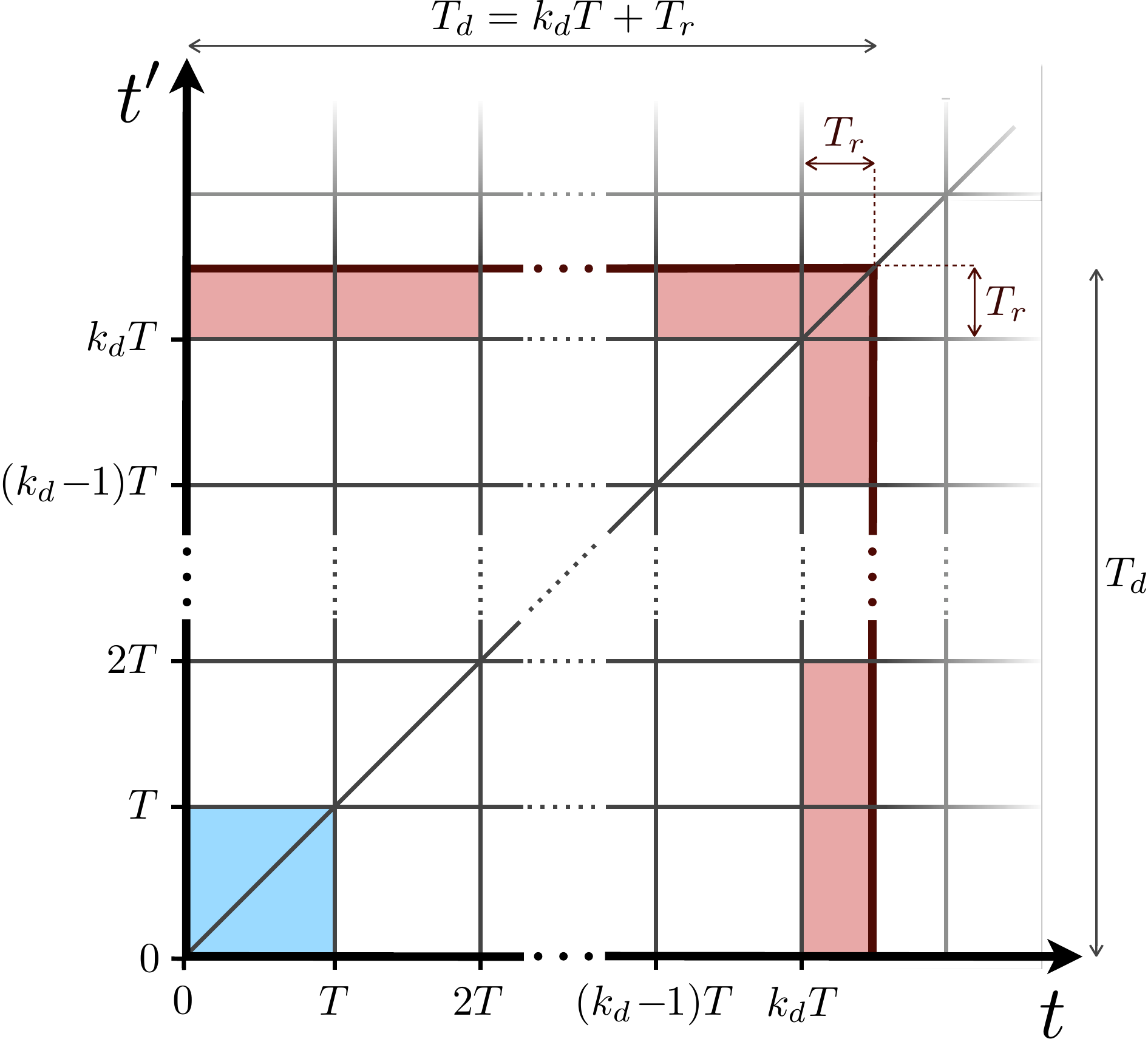}
\caption{Integration domain for the spectral densities. \label{Fig:Chessboard}}
\end{figure}

Starting from the general expression for the spectral density, Eq.
(\ref{def_Sab}), in this appendix we make the derivations required
to turn it into the simplified expression (\ref{Sab_simplified})
provided in the main text. In order to perform the two-time integral
(\ref{def_Sab}) we split the integration domain $\left[0,T_{d}\right]\times\left[0,T_{d}\right]$
into intervals of duration $T$, obtaining a kind of chessboard as
shown in Fig. \ref{Fig:Chessboard}. We denote by $k_{d}$ the number
of full periods contained in the detection interval, which is the
common number of squares along the horizontal and the vertical directions
of the chessboard, and by $T_{r}$ the remainder ($T_{d}=k_{d}T+T_{r}$,
with $k_{d}>0$ and $0\leq T_{r}<T$), which is the width of the red
boundaries in the figure. According to this, we decompose the integral
(\ref{def_Sab}), using Eq. (\ref{Cab}), as 
\begin{equation}
\mathcal{S}_{\alpha\beta}(\omega)=\frac{\Upsilon(\mu_{\alpha}+\mu_{\beta})}{T_{d}}\left[\sum_{\ell=0}^{k_{d}-1}\sum_{m=0}^{k_{d}-1}I_{\alpha\beta}^{(\ell,m)}(\omega)+R_{\alpha\beta}(\omega)\right],\label{Sab}
\end{equation}
where the generic integral 
\begin{align}\label{Ilm_def}
I_{\alpha\beta}^{(\ell,m)}(\omega)&\coloneqq\int_{\ell T}^{(\ell+1)T}dt\int_{mT}^{(m+1)T}dt'P_{\alpha}(t)P_{\beta}(t')
\\
&\hspace{3cm}\times\overline{C}_{\alpha\beta}(t,t')e^{i\omega(t-t')},\nonumber
\end{align}
extends over the square whose lower-left corner seats at $(t=\ell T,t'=mT)$,
and the remainder reads
\begin{align}\label{Rab_def}
R_{\alpha\beta}(\omega)&\coloneqq\sum_{m=0}^{k_{d}-1}H_{\alpha\beta}^{(k_{d},m)}(\omega)
\\
&\hspace{1cm}+\sum_{\ell=0}^{k_{d}-1}H_{\alpha\beta}^{(\ell,k_{d})}(\omega)+H_{\alpha\beta}^{(k_{d},k_{d})}(\omega),\nonumber
\end{align}
where
\begin{subequations}
\begin{align}
H_{\alpha\beta}^{(k_{d},m)}(\omega) & \coloneqq\int_{k_{d}T}^{k_{d}T+T_{r}}\hspace{-2mm}dt\int_{mT}^{(m+1)T}\hspace{-2mm}dt'P_{\alpha}(t)P_{\beta}(t')
\\
&\hspace{3.5cm}\times\overline{C}_{\alpha\beta}(t,t')e^{i\omega(t-t')},\nonumber
\\
H_{\alpha\beta}^{(\ell,k_{d})}(\omega) & \coloneqq\int_{\ell T}^{(\ell+1)T}\hspace{-2mm}dt\int_{k_{d}T}^{k_{d}T+T_{r}}\hspace{-2mm}dt'P_{\alpha}(t)P_{\beta}(t')
\\
&\hspace{3.5cm}\times\overline{C}_{\alpha\beta}(t,t')e^{i\omega(t-t')},\nonumber
\\
H_{\alpha\beta}^{(k_{d},k_{d})}(\omega) & \coloneqq\int_{k_{d}T}^{k_{d}T+T_{r}}\hspace{-2mm}dt\int_{k_{d}T}^{k_{d}T+T_{r}}\hspace{-2mm}dt'P_{\alpha}(t)P_{\beta}(t')\label{Jkdkd}
\\
&\hspace{3.5cm}\times\overline{C}_{\alpha\beta}(t,t')e^{i\omega(t-t')},\nonumber
\end{align}
\end{subequations} are integrals extending over the incomplete squares
at the red boundary of the chessboard in Fig. \ref{Fig:Chessboard},
and we keep the same convention on the upper indices as with the $I$
integrals.

As we show next, the key point is that any of the above integrals, $I$ or $H$, can be algebraically related to integrals defined over the $[0,T]\times[0,T]$ domain, corresponding to the blue square in Fig. \ref{Fig:Chessboard}. Consider first the integrals $I_{\alpha\beta}^{(\ell,m)}(\omega)$ of Eq. (\ref{Ilm_def}). Because $\overline{C}_{\alpha\beta}(t,t^\prime)$ takes on different expressions depending on whether $t'<t$ or $t<t'$,
see Eq. (\ref{overCab}), we must distinguish between integration domains that are above, below, or along the chessboard's diagonal (also represented in Fig. \ref{Fig:Chessboard}). We then distinguish between
integrals with $\ell=m$, and integrals with $\ell>m$ and $\ell<m$, which we will denote respectively as $I_{\alpha\beta}^{(\ell>m)}(\omega)$ and $I_{\alpha\beta}^{(\ell<m)}(\omega)$. When $\ell>m$, then $t'<t$, hence the argument of the noise correlation $\Gamma_{\alpha\beta}$ in Eq. (\ref{overCab}) is $t'\bmod T$, while if $\ell<m$, then $t<t'$, hence the argument is $t\bmod T$. Performing the variable change
$t\rightarrow{t-\ell T}$ and $t'\rightarrow{t'-mT}$ in the integrals, using Eq. (\ref{overCab}), and recalling the assumed $T$-periodicity of the functions $P_{\alpha}$, we get
\begin{subequations}\label{Ilmml}
\begin{align}
I_{\alpha\beta}^{(\ell>m)}(\omega) & =e^{(\ell-m)(\mu_{\alpha}+i\omega)T}I_{\alpha\beta}^{\searrow}(\omega),\\
I_{\alpha\beta}^{(\ell<m)}(\omega) & =e^{(m-\ell)(\mu_{\beta}-i\omega)T}I_{\alpha\beta}^{\nwarrow}(\omega),
\end{align}
\end{subequations}
where
\begin{subequations} 
\begin{align}
I_{\alpha\beta}^{\searrow}(\omega) & \coloneqq\int_{0}^{T}dtP_{\alpha}(t)e^{(\mu_{\alpha}+i\omega)t}
\\
&\hspace{1cm}\times\int_{0}^{T}dt'P_{\beta}(t')\Gamma_{\alpha\beta}(t')e^{-(\mu_{\alpha}+i\omega)t'},\nonumber\label{Iarrow}
\\
I_{\alpha\beta}^{\nwarrow}(\omega) & \coloneqq\int_{0}^{T}dt'P_{\beta}(t')e^{(\mu_{\beta}-i\omega)t'}
\\
&\hspace{1cm}\times\int_{0}^{T}dtP_{\alpha}(t)\Gamma_{\alpha\beta}(t)e^{-(\mu_{\beta}-i\omega)t}.\nonumber
\end{align}
\end{subequations}Note that the ``$\bmod\,T$'' operator has disappeared,
as now integrals extend along $t,t'\in[0,T]$. As for the integrals
$I_{\alpha\beta}^{(\ell,\ell)}(\omega)$, we proceed along the previous
lines, just considering that the argument of $\Gamma_{\alpha\beta}$ in
Eq. (\ref{overCab}) is $t'\bmod T$ in the lower-right half of any
diagonal square (which we denote by ``$\text{\large\ensuremath{\lrcorner}}$''
in the following), while it is $t\bmod T$ in the upper-left one (which
we denote by ``$\rotatebox[origin=c]{180}{\text{\large\ensuremath{\lrcorner}}}$'').
Performing the variable change $t\rightarrow{t-\ell T}$ and $t'\rightarrow{t'-\ell T}$,
we then easily find that all integrals $I_{\alpha\beta}^{(\ell,\ell)}(\omega)$
have the same value,  
\begin{align}\label{Ill}
I_{\alpha\beta}^{(\ell,\ell)}(\omega)=I_{\alpha\beta}^{(0,0)}(\omega)=I_{\alpha\beta}^{\text{\large\ensuremath{\lrcorner}}}(\omega)+I_{\alpha\beta}^{\,\rotatebox[origin=c]{180}{\text{\large\ensuremath{\lrcorner}}}}(\omega),\forall\ell,
\end{align}
where
\begin{subequations}\label{Ilrul}
\begin{align}
I_{\alpha\beta}^{\text{\large\ensuremath{\lrcorner}}}(\omega)\coloneqq & \int_{0}^{T}dt'P_{\beta}(t')\Gamma_{\alpha\beta}(t')e^{-(\mu_{\alpha}+i\omega)t'}
\\
&\hspace{1cm}\times\int_{t'}^{T}dtP_{\alpha}(t)e^{(\mu_{\alpha}+i\omega)t},\nonumber
\\
I_{\alpha\beta}^{\,\rotatebox[origin=c]{180}{\text{\large\ensuremath{\lrcorner}}}}(\omega)\coloneqq & \int_{0}^{T}dtP_{\alpha}(t)\Gamma_{\alpha\beta}(t)e^{-(\mu_{\beta}-i\omega)t}
\\
&\hspace{1cm}\times\int_{t}^{T}dt'P_{\beta}(t')e^{(\mu_{\beta}-i\omega)t'}.\nonumber
\end{align}
\end{subequations}It is interesting to note that when the noise correlation
matrix $\mathcal{G}$ is symmetric, so that $\Gamma_{\alpha\beta}(t)=\Gamma_{\beta\alpha}(t)$,
these integrals satisfy the property $I_{\alpha\beta}^{\text{\large\ensuremath{\lrcorner}}}(\omega)=I_{\beta\alpha}^{\,\rotatebox[origin=c]{180}{\text{\large\ensuremath{\lrcorner}}}}(-\omega)$.

With all previous results we can finally give a compact expression
for the spectral density $S_{\alpha\beta}(\omega)$ defined in Eq.
(\ref{def_Sab}). Substituting Eqs. (\ref{Ilmml}) and (\ref{Ill})
into Eq. (\ref{Sab}), and performing the summations we obtain 
\begin{align}\label{genSab}
&\mathcal{S}_{\alpha\beta}(\omega)=\frac{\Upsilon(\mu_{\alpha}+\mu_{\beta})}{T_{d}/k_{d}}\Bigg[\frac{1}{k_{d}}R_{\alpha\beta}(\omega)+I_{\alpha\beta}^{(0,0)}(\omega)
\\
&\hspace{-1mm}+\varepsilon_{\alpha}(\omega)\Upsilon(\mu_{\alpha}+i\omega)I_{\alpha\beta}^{\searrow}(\omega)+\varepsilon_{\beta}(-\omega)\Upsilon(\mu_{\beta}-i\omega)I_{\alpha\beta}^{\nwarrow}(\omega)\Bigg],\nonumber
\end{align}
where $\Upsilon(x)$ was defined in Eq. (\ref{Yota}), and we have
introduced the auxiliary function 
\begin{align}
\varepsilon_{\alpha}(\omega)&\coloneqq\frac{1}{\Upsilon(\mu_{\alpha}+i\omega)k_{d}}\sum_{\ell=1}^{k_{d}-1}\sum_{m=0}^{\ell-1}e^{(\ell-m)(\mu_{\alpha}+i\omega)T}
\\
&=1-\frac{1}{k_{d}}\frac{1-e^{(\mu_{\alpha}+i\omega)k_{d}T}}{1-e^{(\mu_{\alpha}+i\omega)T}}.\nonumber\label{varepsilon}
\end{align}
Note that when the detection time $T_{d}$ contains very many periods
$T$, i.e. when $k_{d}\to\infty$, the general expression (\ref{genSab})
simplifies, since $\varepsilon_{\alpha}(\omega)\rightarrow1$, $T_{d}/k_{d}\to T$,
and the contribution of the remainder $R_{\alpha\beta}$ becomes negligible.
Thus, in this this limit we obtain exactly the form that we presented
in the main text, Eq. (\ref{Sab_simplified}). Otherwise, the reminder
needs to be evaluated, and for that it is useful to have an expression
referred only to the first period. In order to do this, we simply
proceed in the same manner as we did for the determination of the
integrals $I_{\alpha\beta}^{(\ell,m)}$, now taking into account that
$T_{r}<T$. The remainder defined in Eq. (\ref{Rab_def}) can be reduced,
after working out the summations, to 
\begin{align}
&\hspace{-2mm}R_{\alpha\beta}(\omega)=H_{\alpha\beta}^{(0,0)}+\Upsilon(\mu_{\alpha}+i\omega)\hspace{-1mm}\left(1-e^{(\mu_{\alpha}+i\omega)k_{d}T}\right)\hspace{-1mm}H_{\alpha\beta}^{\searrow}(\omega)\nonumber
\\
&\hspace{1cm}+\Upsilon(\mu_{\beta}-i\omega)\left(1-e^{(\mu_{\beta}-i\omega)k_{d}T}\right)H_{\alpha\beta}^{\nwarrow}(\omega),
\end{align}
where the integral $H_{\alpha\beta}^{(0,0)}$ has been defined as
$H_{\alpha\beta}^{(k_{d},k_{d})}$ in Eq. (\ref{Jkdkd}), setting
$k_{d}\to0$. Hence it formally coincides with $I_{\alpha\beta}^{(0,0)}$ in
Eq. (\ref{Ilm_def}), with the substitution $T\to T_{r}$ in Eqs.
(\ref{Ilrul}), and accordingly it is given by Eq. (\ref{Ill}) with
the latter substitution. We have also defined the following integrals
\begin{subequations}\label{Jarrow} 
\begin{align}
H_{\alpha\beta}^{\searrow}(\omega) & \coloneqq\int_{0}^{T_{r}}dtP_{\alpha}(t)e^{(\mu_{\alpha}+i\omega)t}
\\
&\hspace{1cm}\times\int_{0}^{T}dt'P_{\beta}(t')\Gamma_{\alpha\beta}(t')e^{-(\mu_{\alpha}+i\omega)t'},\nonumber
\\
H_{\alpha\beta}^{\nwarrow}(\omega) & \coloneqq\int_{0}^{T_{r}}dt'P_{\beta}(t')e^{(\mu_{\beta}-i\omega)t'}
\\
&\hspace{1cm}\times\int_{0}^{T}dtP_{\alpha}(t)\Gamma_{\alpha\beta}(t)e^{-(\mu_{\beta}-i\omega)t}.\nonumber
\end{align}
\end{subequations}

\section{Working out cross-correlations and cross-spectral densities with
the noise \label{Appendix_CrossCorrNoise}}

In this appendix we explain how we have dealt with the cross-correlations
between the projections and the noise, as well as with the corresponding
spectral density, in order to find the simplified expressions of Eqs.
(\ref{def_Ccross}) and (\ref{def_Scross}). Regarding the correlation
functions, these are immediately found by using the solution (\ref{sol_c})
and the form of the projected noise (\ref{ProjNoises}). In particular,
we get
\begin{align}
\langle\hat{c}_{\alpha}(t)\hat{\xi}_{\beta}(t')\rangle&=\int_{-\infty}^{t}dt_{1}e^{\mu_{\alpha}(t-t_{1})}
\\
&\hspace{8mm}\times\sum_{\sigma}\left[\mathcal{K}^{-1}(t_{1})\mathcal{B}(t_{1})\right]_{\alpha\sigma}\underbrace{\langle\hat{\xi}_{\sigma}(t_{1})\hat{\xi}_{\beta}(t')\rangle}_{\mathcal{G}_{\sigma\beta}\delta(t_{1}-t')}\nonumber
\\
&=\begin{cases}
e^{\mu_{\alpha}(t-t')}\left[\mathcal{K}^{-1}(t')\mathcal{B}(t')\mathcal{G}\right]_{\alpha\beta}, & t'\leq t
\\
0, & t<t'
\end{cases},\nonumber
\end{align}
which is precisely the expression (\ref{def_Ccross}) that we provide
in the main text. Proceeding in the same way, one finds the expression
for $\langle\hat{\xi}_{\alpha}(t)\hat{c}_{\beta}(t)\rangle$ shown
in (\ref{def_Ccross}).

As for the spectral density associated to $\mathcal{C}_{\alpha\beta}^{(c\xi)}(t,t')=\langle\hat{c}_{\alpha}(t)\hat{\xi}_{\beta}(t')\rangle$,
we simply need to note that this cross-correlation is zero in the
upper triangular region of the integration domain of Fig. \ref{Fig:Chessboard},
while in the lower triangular it has the same form as $\mathcal{C}_{\alpha\beta}(t,t')=\langle\hat{c}_{\alpha}(t)\hat{c}_{\beta}(t')\rangle$
in Eq. (\ref{Cab-1}), just replacing $\Upsilon(\mu_{\alpha}+\mu_{\beta})\Gamma_{\alpha\beta}(t'\bmod{T})$
by $\chi_{\alpha\beta}^{(c\xi)}(t'\bmod{T})=\left[\mathcal{K}^{-1}(t'\bmod{T})\mathcal{B}(t'\bmod{T})\mathcal{G}\right]_{\alpha\beta}$,
where we have used the periodicity of $\mathcal{K}(t')$ and $\mathcal{B}(t')$.
Hence, it is clear that using the same derivations as in the previous
appendix, in particular the ones turning Eq. (\ref{Sab}) into Eq.
(\ref{genSab}), one obtains the spectral density introduced in (\ref{def_Scross})
after taking the $k_{d}\rightarrow\infty$ limit. A similar argument
applies to the spectral density of the other cross-correlation $\mathcal{C}_{\alpha\beta}^{(\xi c)}(t,t')=\langle\hat{\xi}_{\alpha}(t)\hat{c}_{\beta}(t')\rangle$,
just noting that this one is zero in the lower triangular region of the integration
domain of Fig. \ref{Fig:Chessboard}.

\end{document}